\documentclass[aps,prl,twocolumn,longbibliography,noeprint,floatfix,superscriptaddress]{revtex4-2}

\usepackage{amsmath, amssymb}
\usepackage{mathtools}
\usepackage{lipsum}
\usepackage{times}
\usepackage{graphicx}
\usepackage{wasysym}
\usepackage{bm}
\usepackage{hyperref}
\usepackage{xspace}
\usepackage{siunitx}
\usepackage[usenames,dvipsnames]{xcolor}
\usepackage{soul}
\usepackage{centernot}
\usepackage{braket}
\usepackage{microtype}

\usepackage{titlesec}

\titlespacing\section{0pt}{12pt plus 4pt minus 2pt}{0pt plus 2pt minus 2pt}
\titlespacing\subsection{0pt}{12pt plus 2pt minus 2pt}{11pt}
\titlespacing\subsubsection{0pt}{12pt plus 4pt minus 2pt}{0pt plus 2pt minus 2pt}

\definecolor{myColor2}{rgb}{0.02,0.12,0.3}

\definecolor{myColor}{rgb}{0.02,0.12,0.7}
\definecolor{myciteColor}{rgb}{0.39,0.7,0.89}
\hypersetup{colorlinks=true, linkcolor=myColor, filecolor=myColor, urlcolor=myColor, citecolor=myColor, urlcolor=myColor}
\graphicspath{{./Figures/}}
\DeclareSIUnit{\nK}{\nano\kelvin}
\DeclareSIUnit{\aB}{\emph{a}_0}
\DeclareSIUnit{\G}{G}

\newcommand{\kB}{k_{\text{B}}}

\newcommand{\tref}{t_{0}}
\newcommand{\Tbkt}{T_{\text{BKT}}}
\newcommand{\focus}{\ensuremath{f_{0}}\xspace}

\newcommand{\tuni}{\ensuremath{t_{\text{uni}}}\xspace}
\newcommand{\tstar}{\ensuremath{t^*}\xspace}

\begin{document}

%\preprint{APS/123-QED}

\title{
Universal Coarsening in a Homogeneous Two-Dimensional Bose Gas
}
\author{Martin Gazo}
\email{mg816@cam.ac.uk}
\affiliation{Cavendish Laboratory, University of Cambridge, J. J. Thomson Avenue, Cambridge CB3 0HE, United Kingdom}
\author{Andrey Karailiev}
\affiliation{Cavendish Laboratory, University of Cambridge, J. J. Thomson Avenue, Cambridge CB3 0HE, United Kingdom}
\author{Tanish Satoor}
\affiliation{Cavendish Laboratory, University of Cambridge, J. J. Thomson Avenue, Cambridge CB3 0HE, United Kingdom}
\author{Christoph Eigen}
\affiliation{Cavendish Laboratory, University of Cambridge, J. J. Thomson Avenue, Cambridge CB3 0HE, United Kingdom}
\author{Maciej Ga\l ka}
\affiliation{Cavendish Laboratory, University of Cambridge, J. J. Thomson Avenue, Cambridge CB3 0HE, United Kingdom}
\affiliation{Physikalisches Institut der Universität Heidelberg,
Im Neuenheimer Feld 226, 69120 Heidelberg, Germany}
\author{Zoran Hadzibabic}
\affiliation{Cavendish Laboratory, University of Cambridge, J. J. Thomson Avenue, Cambridge CB3 0HE, United Kingdom}

\begin{abstract}

Coarsening of an isolated far-from-equilibrium quantum system is a paradigmatic many-body phenomenon, relevant from subnuclear to cosmological lengthscales, and predicted to feature universal dynamic scaling. Here, we observe universal scaling in the coarsening of a homogeneous two-dimensional Bose gas, with exponents that match analytical predictions. For different initial states, we reveal universal scaling in the experimentally accessible finite-time dynamics by elucidating and accounting for the initial-state-dependent prescaling effects. The methods we introduce establish direct comparison between cold-atom experiments and non-equilibrium field theory, and are applicable to any study of universality far from equilibrium.

\end{abstract}
\maketitle

%%%%%%%%%%%%%%%%%%%%%%%%%%%%%%%%%%%%%%%%%%%%%%%%%%%%%%%%%%%%%%%%%%%%%

\begin{figure}[t]
\centering
\includegraphics[width=\columnwidth]{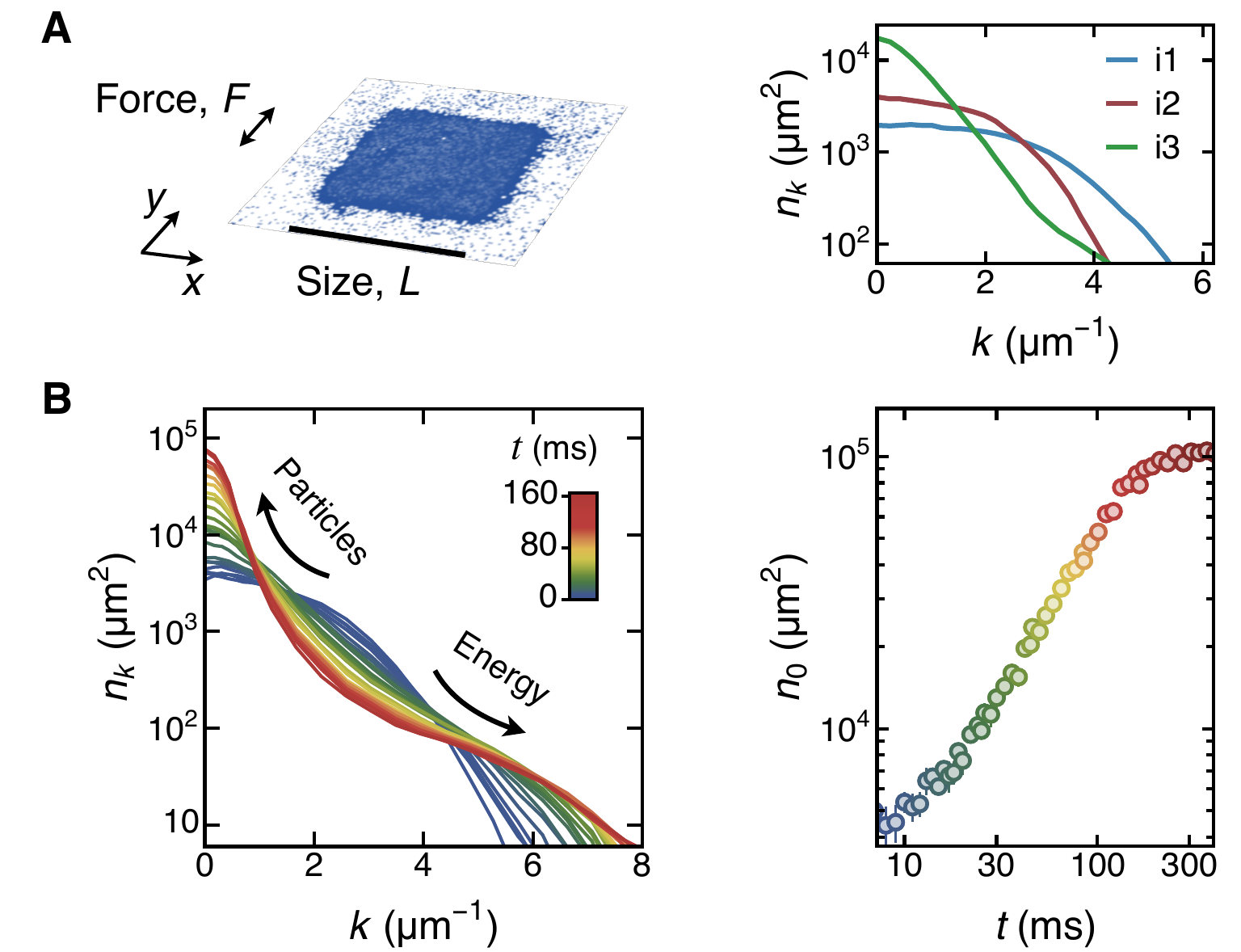}
\caption{\textbf{Bidirectional relaxation in a box-trapped 2D Bose gas.} (\textbf{A})~Starting with a quasi-pure condensate in a box of size $L=\SI{50}{\um}$, we temporarily turn off the interatomic interactions and drive the gas with an oscillating force $F$ to engineer different far-from-equilibrium momentum distributions i1--i3. (\textbf{B}) After we stop the driving and turn on interactions (at $t=0$), the gas exhibits bidirectional relaxation, shown here for i2: most particles flow to the IR (low-$k$ modes), while the net energy flow is to the UV (high $k$). The right panel shows $n_0(t) = n_k(k=0, t)$, which characterizes the growth of the condensate; for $t \gtrsim \SI{150}{\ms}$ the observed growth of $n_0$ is limited by our $k$-space resolution. All measurements are repeated about 30 times and the error bars show standard error of the mean.
}
 \label{fig:1}
\end{figure}

%%%%%%%%%%%%%%%%%%%%%%%%%%%%

Understanding the dynamics of order formation in many-body systems is a long-standing problem in contexts including critical phenomena~\cite{Hohenberg:1977}, driven systems~\cite{Cross:1993}, and active matter~\cite{Ramaswamy:2010,Bowick:2022}.
Of particular interest are general ordering principles that are independent of the microscopic details of a system. 
A prime example of this is coarsening,  the emergence of order characterized by a single lengthscale that grows in time~\cite{Bray:2002, Cugliandolo:2015}.
In systems that can exchange energy with the environment, coarsening is intuitively
linked with cooling, most dramatically through a phase transition. However, the problem is more intricate if a far-from-equilibrium system is isolated and relaxes towards equilibrium only under the influence of internal interactions~\cite{Damle:1996,Berges:2008,Polkovnikov:2011,Mikheev:2023}. 
In this case coarsening commonly involves bidirectional dynamics in momentum space, with most excitations flowing towards lower wavenumbers $k$, but the conserved energy flowing to high $k$.

The framework of non-thermal fixed points (NTFPs)~\cite{Berges:2008,Berges:2009,Mikheev:2023} aims to provide a unifying picture of relaxation in isolated far-from-equilibrium systems, including the early universe undergoing reheating~\cite{Micha:2004}, quark-gluon plasma in heavy-ion collisions~\cite{Berges:2014b}, quantum magnets~\cite{Bhattacharyya:2020}, and ultracold atomic gases~\cite{Nowak:2012,PineiroOrioli:2015,Berges:2015,Karl:2017, Chantesana:2019,Chatrchyan:2021,Groszek:2021,Gresista:2022}. 
In all these cases, relaxation is expected to feature self-similar dynamic scaling; this spatiotemporal analogue of the spatial scale-invariance in equilibrium systems near critical points~\cite{Chaikin:1995} could allow formulation of universality classes for far-from-equilibrium dynamics.

%%%%%%%%%%%%%%%%%%%%%%%%%%%%%%%%%%%%%%%%%%%
\begin{figure*}[t]
\centering
\includegraphics[width=\textwidth]{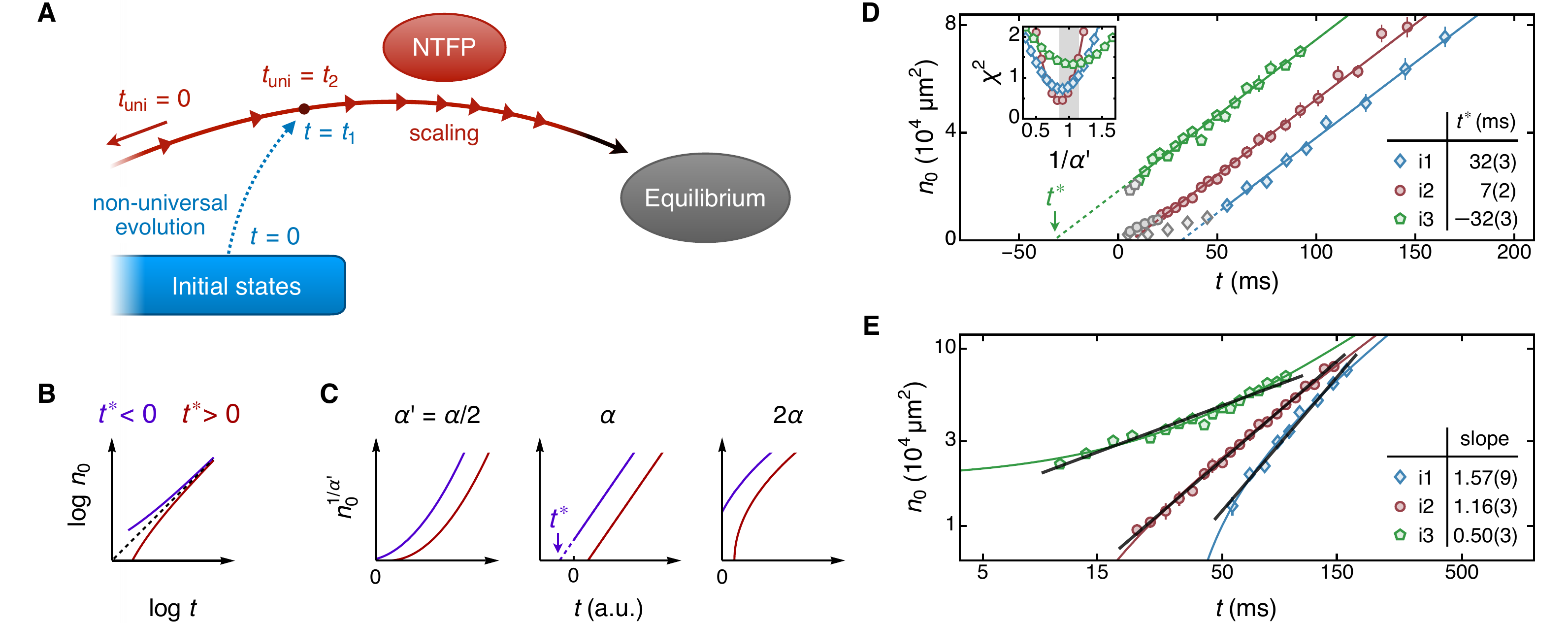}
\caption{
\textbf{Non-universal dynamics, prescaling, and the universal-scaling exponents.}
(\textbf{A}) A system starting in a generic initial state takes some non-universal time $t_1$ to join the scaling trajectory (red) associated with a non-thermal fixed point (NTFP). At $t_1$, its state is the same as if it had always been on this trajectory and evolved for some $t_2 \neq t_1$. At $t> t_1$, it exhibits scaling with respect to the universal-clock time $\tuni = t - \tstar$, where $\tstar = t_1-t_2$ can be positive or negative.
(\textbf{B}) $n_0 \propto (t - \tstar)^{\alpha}$ shows `prescaling', with a flowing exponent ${\rm d} \ln(n_0)/{\rm d} \ln(t) = \alpha/(1-\tstar\!/t)$ that asymptotically approaches $\alpha$. 
(\textbf{C}) One can deduce $\alpha$ from finite-$t$ data using the fact that $n_0^{1/\alpha'}\!(t)$ is linear only for $\alpha' = \alpha$.
(\textbf{D})
Analysis of our i1--i3 data. The main panel shows that the predicted $\alpha = 1$ gives straight lines that differ just in the intercepts $\tstar$, and the inset shows $\chi^2(\alpha')$ for linear fits of $n_0^{1/\alpha'}\!$. The gray symbols indicate early times excluded from the analysis, and the shading in the inset indicates $1/\alpha' = 1.00\pm 0.15$.
(\textbf{E}) Log-log plots of $n_0(t)$ show different prescaling for each initial state. The converging colored lines, with slopes $1$ for $t\gg|\tstar|$, are the same as the straight ones in (D), while the na{\"i}ve power-law fits shown in black have slopes that vary between $0.5$ and $1.6$.
}
 \label{fig:2}
\end{figure*}
%%%%%%%%%%%%%%%%%%%%%%%%%%%%%% 

In experiments with highly controllable Bose gases, self-similar dynamics akin to the NTFP predictions have been observed in a range of scenarios, including different dimensionalities and relaxation of both particle and spin distributions~\cite{Prufer:2018, Erne:2018, Glidden:2021, Orozco:2022,Lannig:2023,Huh:2024}. However, many experimental and theoretical questions regarding the values of the scaling exponents remain open. A key challenge is that, for a generic initial state, universal scaling can be directly observed only for very long evolution times~\cite{Kendon:2001,Mazeliauskas:2019,Schmied:2019,Mikheev:2023,Heller:2023}, which are difficult to access experimentally.

Here, we study coarsening in a homogeneous two-dimensional (2D) Bose gas~\cite{Chomaz:2015,Navon:2021}, by engineering different far-from-equilibrium initial states and measuring the momentum distributions $n_k(k)$ by matter-wave focusing~\cite{Tung:2010,SI}. Crucially, by elucidating and accounting for the non-universal effects of initial conditions, we reveal the universal long-time scaling in finite-range experimental data, introducing methods applicable to any quantitative study of universality far from equilibrium.
We find that the low-$k$ (IR) coarsening is characterized by the theoretically predicted dynamical exponent $z=2$~\cite{Chantesana:2019,Gresista:2022} (see also~\cite{Bray:2002,Damle:1996,Karl:2017,Groszek:2021,Proukakis:2023}), and the form of the self-similarly evolving $n_k$ matches an analytical field-theory prediction~\cite{Chantesana:2019}, while the high-$k$ (UV) energy dynamics corresponds to weak four-wave turbulence~\cite{Dyachenko:1992}.
\subsection*{The experiment: bidirectional relaxation}

We start with a quasi-pure interacting 2D condensate of $7\times 10^4$ atoms of $^{39}$K in the lowest hyperfine state, confined in a square box trap of size $L = \SI{50}{\micro\metre}$~\cite{Christodoulou:2021}. The interactions in the gas, characterized by the scattering length $a$, are tuneable via the magnetic Feshbach resonance at $\SI{402.7}{\G}$~\cite{Etrych:2023}. To prepare our far-from-equilibrium initial states, we temporarily turn off the interactions ($a \rightarrow 0$) and shake the gas with a spatially uniform oscillating force $F$ (see Fig.~\ref{fig:1}{A}). This destroys the condensate and, as previously studied in 3D~\cite{Martirosyan:2023,YZhang:2023}, results in an isotropic highly nonthermal $n_k$ distribution. 
After preparing one of the three different initial states i1--i3 shown in Fig.~\ref{fig:1}A, we stop the shaking, reinstate the interactions ($a\rightarrow 30\,a_0$, where $a_0$ is the Bohr radius), and let the gas relax. 
The states i1--i3 do not have a defined temperature, but $E = \int \varepsilon(k) \, {\rm d}k$, where $\varepsilon = 2\pi \hbar^2 k^3 n_k/(2m)$ and $m$ is the atom mass, gives the total energy. We get $E/\kB = 4.1(3)\,$mK, $2.2(3)\,$mK, and $1.0(3)\,$mK, for i1--i3 respectively; in all cases $E$ is sufficiently low for a condensate to emerge during relaxation~\footnote{For our trap parameters,  $a=30\,a_0$ corresponds to the dimensionless 2D coupling strength $\tilde{g} = 0.026$ and the Berezinskii-Kosterlitz-Thouless critical temperature is $\Tbkt = \SI{230}{\nK}$~\cite{Hadzibabic:2011}}.

In Fig.~\ref{fig:1}{B} we show, for i2, the bidirectional dynamics of the full $n_k$ distribution (left panel) and the growth of $n_0 = n_k(k=0)$, which characterizes the growth of the condensate (right panel). Here $t=0$ corresponds to the time when the interactions are switched on and relaxation starts. 

%%%%%%%%%%%%%%%%%%%%%%%%%%%%%%%%%%%%%%%%%%%

\begin{figure*}[t!]
    \centering
\includegraphics[width=\textwidth]{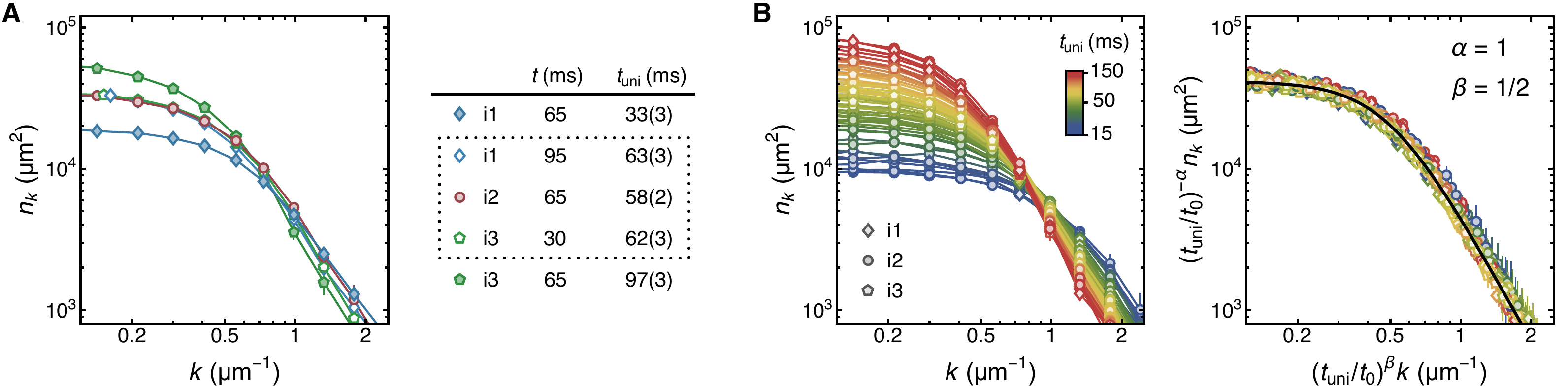}
\caption{ \textbf{Universal coarsening.} (\textbf{A}) For i1--i3, the IR distributions after the same evolution time $t$ (here $65\,$ms) are different, but $n_k$ at the same universal-clock time $\tuni = t - \tstar$ (here $\approx 60\,$ms) are the same; the $\tstar$ values for i1--i3 are those independently determined in Fig.~\ref{fig:2}{D}. (\textbf{B}) The $n_k$ curves for all three initial states collapse onto a universal curve (right panel) according to Eq.~(\ref{eq:1}) with $t \rightarrow \tuni$, $\alpha = 1$, and $\beta = 1/2$, without any free parameters; $\tref = \SI{80}{\ms}$ is an arbitrary reference time. The solid black line shows a fit to the collapsed data with $A/(1+(k/k_0)^3)$.
}
 \label{fig:3}
 \vspace{-1mm}
\end{figure*}
%%%%%%%%%%%%%%%%%%%%%%%%%%%%%%%%%

\subsection*{Dynamic scaling theory
}

According to the dynamic scaling hypothesis, the evolution of $n_k$ is given by
\begin{equation}
    n_k (k,t) = \big{(}\tfrac{t}{\tref}\big{)}^\alpha \ell_{0}^{d} \, f\big{(}\big{(} \tfrac{t}{\tref}\big{)}^\beta k \ell_0 \big{)}  \,,
\label{eq:1}
 \end{equation}
where $\alpha$ and $\beta$ are the scaling exponents, $d$ is the system dimensionality, $f$ is a dimensionless scaling function, and $\ell_0$ is the characteristic lengthscale at an arbitrary reference time $\tref$.
For bidirectional dynamics, such scaling should apply separately in the IR (with $\alpha, \beta >0$) and the UV (with $\alpha, \beta <~0$), with the normalization of $f$ depending on the value of the transport-conserved quantity (particle number or energy).

For the IR dynamics in our 2D gas, analytical theory for compressible (wave) excitations~\cite{Chantesana:2019,Mikheev:2023} predicts $\beta = 1/z = 1/2$ and $\alpha=d\beta =1$, with the ratio \mbox{$\alpha/\beta = 2$} reflecting particle-conserving transport; coarsening is then captured by the scaling of the characteristic lengthscale $\ell(t) = \left( {t}/{\tref}\right)^{\beta} \ell_0$.
For the scaling function, predictions from numerical calculations are captured by $f\propto 1/(1+(k/k_0)^\kappa)$~\cite{Gresista:2022} with $k_0 \propto 1/\ell_0$ and analytically predicted exponent $\kappa = 3$ \cite{Chantesana:2019}. In contrast, for generic vortex-dominated initial states, the scaling-function exponent is analytically predicted to be $\kappa=4$~\cite{Bray:2002,Nowak:2012}, while numerical simulations predict essentially the same $\alpha$ and $\beta$ as for compressible excitations~\cite{Karl:2017}.  

For the UV dynamics in our gas, the prediction based on the theory of weak four-wave turbulence~\cite{Dyachenko:1992, Nazarenko:2011} is $\beta=-1/6$ and $\alpha= (d+2)\beta = -2/3$~\footnote{The corresponding prediction for weak three-wave turbulence, $\beta = -1/4$ and $\alpha=-1$, assumes a time-independent $n_0$ and is not applicable here. If the UV dynamics in our experiments were dominated by three-wave interactions, they would have to be even `faster' (with larger $|\beta|$), due to the growing $n_0$.}, with $\alpha/\beta =4$ reflecting energy-conserving transport. In this case, the theory does not predict the scaling function $f$.

Crucially, for a far-from-equilibrium system to display universal scaling dynamics, it must first `forget' its specific initial state (but not yet approach equilibrium).

\subsection*{Stages of relaxation: from non-universal to universal}

Starting with the IR dynamics, specifically the growth of $n_0$, for which Eq.~(\ref{eq:1}) reduces to  
\begin{equation}
n_0 \propto t^{\alpha}  \propto \ell^d(t)  \,,
\label{eq:2}
\end{equation}
in Fig.~\ref{fig:2} we outline the different stages of relaxation and show how to extract $\alpha$ from finite-$t$ data. For a generic initial state, it first takes some non-universal time $t_1$ for $n_k$ to acquire the scaling form $f$. At $t_1$ the system joins the scaling trajectory shown in red in Fig.~\ref{fig:2}{A}, which follows Eq.~(\ref{eq:1}). However, importantly, if the system had always followed Eq.~(\ref{eq:1}), starting with $n_0=0$ at $t=0$, it would have arrived to the same point on the scaling trajectory at some $t_2$ that in general differs from $t_1$.
\footnotetext[20]{{Given the value of $n_0$ at $t_1$, the universal-clock time $t_2$ must be such that $\Delta t$ later $n_0 = n_0(t_1) [(t_2 + \Delta t)/t_2]^{\alpha}$. For $t = t_1 + \Delta t$, this gives $n_0 (t) = n_0 (t_1) [(t - \tstar)/(t_1 - \tstar)]^{\alpha} \propto (t - \tstar)^{\alpha}$}}%
\footnotetext[21]{Note that a similar mathematical structure appears in the analysis of finite-capacity turbulent cascades~\cite{Nazarenko:2011}, but with $t\rightarrow \tstar\! - t$; in that case $\tstar$ denotes the time when the scaling ends.}%
Defining $\tstar = t_1 - t_2$ and the `universal-clock time' $\tuni = t-\tstar$, time-invariance of the Hamiltonian dictates that from thereon \textls[-25]{$n_0 \!\propto \tuni^{\alpha} = (t-\tstar)^{\alpha}$}~\cite{Note20,Note21}. The system then exhibits `prescaling'~\cite{Mazeliauskas:2019,Schmied:2019, Heller:2023}, with the flowing exponent ${\rm d} \ln(n_0)/{\rm d} \ln(t) = \alpha/(1-\tstar\!/t)$~(see Fig.~\ref{fig:2}{B}). Hence, only for $t\gg |\tstar|$ the physical evolution is given by Eq.~(\ref{eq:1}). 
In essence, by $t_1$ the system forgets what its initial state was, but the memory that it had not always been on the scaling trajectory fades slowly.

Eventually, (pre)scaling breaks down if a system approaches equilibrium, which in a finite size (experimentally relevant) system happens in finite time. 
Our data in Fig.~\ref{fig:1}{B} is suggestive of such behavior; however, note that the maximum $n_0$ we can measure is limited by our $k$-space resolution, so the (pre)scaling regime might be extending further in time than we can observe.

In practice, the regime when $t \gg |\tstar|$ but the system is still far from equilibrium may not be accessible, or even exist.
However, one can deduce $\alpha$ from the finite-$t$ prescaling data ($t>t_1$, but not necessarily $t \gg |\tstar|$), as we show in Fig.~\ref{fig:2}{C} for two different $\tstar$ values: plotting $n_0^{1/{\alpha}'}\!(t)$ for various $\alpha'$ gives a linear plot (with intercept $\tstar$) only for $\alpha' = \alpha$.

In Fig.~\ref{fig:2}{D} we show the results of such analysis for our data taken with initial states i1--i3. The predicted $\alpha = 1$ gives three parallel straight lines, with the differences between the initial states fully captured by the intercepts $\tstar$. In the inset we show that requiring the linearity of $n_0^{1/\alpha}$ for i1--i3 separately gives consistent $\alpha$ values.

For comparison, $n_0(t)$ plots in Fig.~\ref{fig:2}{E} show different prescaling for each initial state. Here, the slopes of na{\"i}ve  power-law fits (black), which approximate the instantaneous values of the flowing prescaling exponents, have values that vary between $0.5$ and $1.6$.
Moreover, for the same data, these non-universal values also depend on how one defines $t=0$; this choice is arbitrary, because any state during evolution can be treated as the initial one for further evolution. In contrast, our extraction of $\alpha$ is independent of this choice, since any arbitrary shift of $t$ is simply absorbed by $\tstar$ (see~\cite{SI}).

%%%%%%%%%%%%%%%%%%%%%%%%%%%%%%%%%%%%%%%%%%%%%%%%%%%
\subsection*{Universal coarsening}

We now turn to the dynamics of the full IR distributions, leveraging the fact that we have deduced the non-universal $\tstar$ values for i1--i3 from the $n_0$ data.
In Fig.~\ref{fig:3}{A} we show five $n_k$ curves, for different initial states and evolution times, which illustrate that distributions corresponding to the same  $\tuni = t-\tstar$ are the same, irrespective of the initial conditions. 

In Fig.~\ref{fig:3}{B} we show all three data sets color-coded according to $\tuni$, and that they all collapse onto the same universal curve when scaled according to Eq.~\eqref{eq:1} with $t \rightarrow \tuni$, $\alpha = 1, \beta = 1/2$, and no free parameters (see also~\cite{SI}). The collapsed data is fitted well by $A/(1+(k/k_0)^\kappa)$ with fixed (analytically predicted) $\kappa = 3$~\cite{Chantesana:2019}; treating $\kappa$ as a free parameter gives $\kappa = 2.9(1)$.

Note that universality requires only the exponents $\alpha$, $\beta$, and $\kappa$ to be the same for different initial states, while we observe that the IR dynamics for i1--i3, measured with respect to $\tuni$, are essentially identical, even though the three states have very different energies~\footnote{This is also hinted at by the fact that the three lines in Fig.~\ref{fig:2}D are parallel.}. This `superuniversality' is restricted to states that have the same value of the transport-conserving quantity, which in the IR is the atom number.

%%%%%%%%%%%%%%%%%%%%%%%%%
\begin{figure*}[t]
\centering

\includegraphics[width=\textwidth]{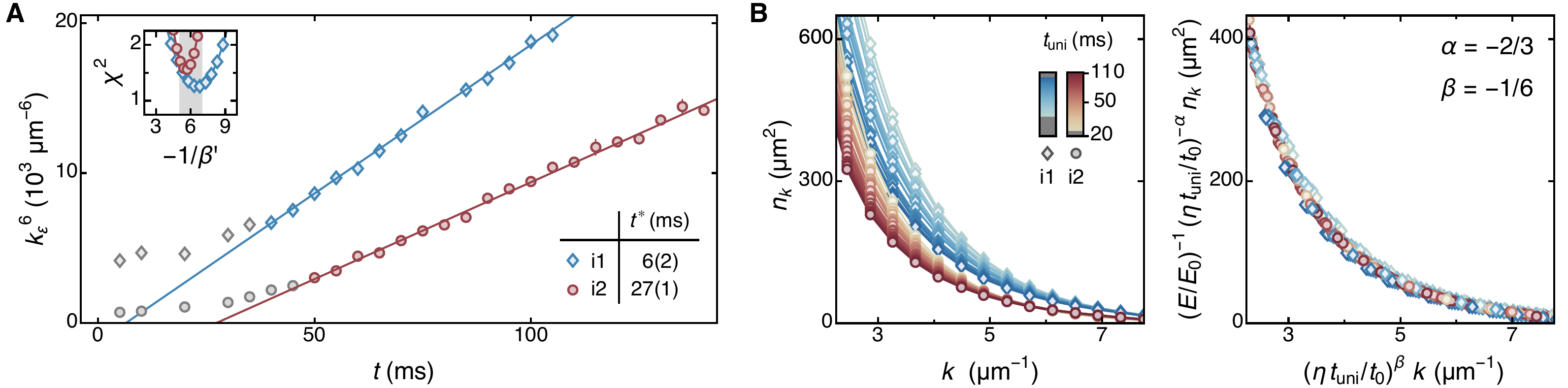}

\caption{ 
\textbf{UV dynamics, for i1 and i2.} (\textbf{A}) Evolution of $k_\varepsilon$, the position of the peak of the energy spectrum, $\varepsilon\propto k^3 n_k$ (see~\cite{SI}). For the predicted $\beta = -1/6$, plotting $k_\varepsilon^{-1/\beta}(t)$ gives straight lines with intercepts $\tstar$; the gray symbols indicate early times excluded from the analysis. The inset shows $\chi^2 (\beta')$ for linear fits of $k_\varepsilon^{-1/\beta'}\!(t)$, with the shading indicating $-1/\beta' = 6\pm1$.
(\textbf{B}) Dynamic scaling of the UV momentum distributions. The i1 and i2 curves labeled by $\tuni = t-\tstar$ (left panel) collapse onto a single universal curve (right panel)  when we: (i) set $\alpha = -2/3$, $\beta = -1/6$, and the arbitrary $\tref = \SI{80}{\ms}$, (ii) account for the different (energy-dependent) clock speeds by setting $\eta ({\rm i2})=1$ and $\eta ({\rm i1})=1.53$, the ratio of the ${\rm d}k_\varepsilon^6/{\rm d}t$ slopes observed in (A), and (iii) normalize $n_k$ by $E$, with $E_0$ arbitrarily set to $E({\rm i2}) = \kB \times 2.2\,$mK.
}
 \label{fig:4}
\end{figure*}
%%%%%%%%%%%%%%%%%%%%%%%%%%%%%%%%%%%%%%%%
\subsection*{UV dynamics}

Finally, we discuss the complementary UV dynamics, for which the ideas from Fig.~\ref{fig:2} also apply. For an arbitrary initial $n_k$, there is no reason for $t_1$ or $\tstar$ to be the same as in the IR, and in the UV there is no fixed $k$ at which $n_k \propto (t-\tstar)^{\alpha}$ for $t>t_1$. However, the peaks of our energy spectra, $\varepsilon (k) \propto k^3 n_k$, provide a characteristic wavenumber $k_\varepsilon (t)$ (see~\cite{SI}), for which 
Eq.~\eqref{eq:1} gives the asymptotic behavior $k_\varepsilon \propto t^{-\beta}$, and more generally for $t>t_1$ one expects:
\begin{equation}
 k_\varepsilon^{-1/\beta} \propto t - \tstar .
\label{eq:3}
\end{equation}
One can thus use the ideas from Fig.~\ref{fig:2}{C} to determine $\beta$ and~$\tstar\!$.

In Fig.~\ref{fig:4} we show our results for initial states i1 and i2; for i3, with the lowest energy $E$, our high-$k$ signal is too weak for a quantitative study. 
In Fig.~\ref{fig:4}{A} we show the linearity of $k_\varepsilon^{-1/\beta}$ (with $\tstar$ intercepts) for the predicted $\beta = -1/6$, and that treating $\beta$ as a free parameter gives consistent values for both states (inset). Here the transport-conserved quantity is $E$, which is not the same for i1 and i2, and we do not observe the superuniversality seen in the IR; 
while $\beta$ is the same, ${\rm d}k_\varepsilon^6/{\rm d}t$ is larger for the higher-energy i1, corresponding to a faster-running $\tuni$ (faster evolution along a scaling trajectory).

In Fig.~\ref{fig:4}{B} we show the dynamic scaling of the UV distributions. Simply setting $t\rightarrow \tuni$, $\alpha = -2/3$, and $\beta = -1/6$ in Eq.~(\ref{eq:1}) collapses the $n_k$ curves shown in the left panel onto two separate curves, with normalizations set by $E({\rm i1})$ and $E({\rm i2})$ (not shown; see~\cite{SI}). Here, in the right panel, we normalize $n_k$ by $E$, and the clock speeds by the observed ${\rm d}k_\varepsilon^6/{\rm d}t$ slopes, which collapses all the data onto a single universal curve.

\subsection*{Conclusions and outlook}

Our experiments provide a comprehensive picture of universal coarsening in a 2D Bose gas dominated by wave excitations, in  parameter-free agreement with theory.
More broadly, we establish methods for direct quantitative comparisons of experiments and analytical field theories of far-from-equilibrium quantum phenomena. Specifically, we show how to extract the theoretically relevant asymptotic long-time evolution from the experimentally accessible finite-time dynamics.
In the future, similar studies with vortex-rich initial states~\cite{Gauthier:2019,Johnstone:2019} could reveal dynamic scaling with anomalous exponents~\cite{Karl:2017,Groszek:2020}, while studies in gases with a supersolid ground state~\cite{Bland:2022} could reveal a fascinating interplay of coarsening and pattern-formation dynamics.
It would also be interesting to extend our work to (miscible) two-component gases, with one component acting as a bath for the other, which would allow studies of the effects of a tuneable openness of the system~\cite{Groszek:2021}.

%%%%%%%%%%%%%%%%
\vspace{1em}

{\bf Acknowledgments.} We thank Gevorg Martirosyan,  Jiří Etrych, Christopher Ho, Yansheng Zhang, Nishant Dogra, Panagiotis Christodoulou, Robert Smith, Thomas Gasenzer, Markus Oberthaler, and Aleksas Mazeliauskas for discussions and comments on the manuscript. This work was supported by EPSRC [Grant No.~EP/P009565/1], ERC (UniFlat), and STFC [Grant No.~ST/T006056/1]. M. Ga{\l}ka acknowledges support from the Germany’s
Excellence Strategy EXC2181/1-390900948 (Heidelberg Excellence Cluster STRUCTURES). Z.H. acknowledges support from the Royal Society Wolfson Fellowship. {\bf Author contributions.} M. Gazo led the data collection and analysis. All authors contributed to the interpretation of the results, with most significant contributions from M. Gazo, M. Ga{\l}ka and Z.H. All authors contributed to the preparation of the manuscript. Z.H. supervised the project.

%

%%%%%%%%%%%%%%%%%%%%%%%%%%%%%%%%%%%%%%%%%%%%%%%%%%%%
%%%%%%%%%%%%%%%%%%%%%%%%%%%%%%%%%%%%%%%%%%%%%%%%%%%%
%%%%%%%%%%%%%%%%%%%%%%%%%%%%%%%%%%%%%%%%%%%%%%%%%%%%

\clearpage
\setcounter{figure}{0} 
\setcounter{equation}{0}

\renewcommand\theequation{S\arabic{equation}} 
\renewcommand\thefigure{S\arabic{figure}}

\titlespacing\section{0pt}{12pt plus 4pt minus 2pt}{4pt}
\titlespacing\subsection{0pt}{12pt}{12pt}

\section{SUPPLEMENTARY MATERIALS}

\vspace{2em}

\phantomsection
\subsection*{Measurements of momentum distributions \texorpdfstring{$n_k$}{nk}}

We use matter-wave focusing~\cite{Tung:2010} to measure $n_k(k)$. Following the relaxation time $t$, we turn off the 2D trap and interactions ($a \rightarrow 0$), and turn on harmonic confinement with a tuneable isotropic in-plane frequency \focus, created by two crossed laser beams. After a quarter of the trap period, $1/(4\focus)$, when the spatial atom distribution reflects $n_k$ at the start of focusing, we transfer a variable fraction of atoms from the $\ket{F=1,m_F=1}$ state to $\ket{F=2,m_F=2}$, and measure the density distribution by absorption imaging on the cycling $\ket{F=2,m_F=2} \rightarrow \ket{F^\prime=3,m_{F^\prime}=3}$ transition. Varying the transferred fraction between 2 and 100\% allows us to measure $n_k$ values over four orders of magnitude.

Generally, a larger \focus minimizes the effects of the focusing-trap anharmonicity and gives better signal-to-noise ratio, making it more suitable for measurements of the high-$k$ (UV) part of the distribution, while a smaller $\focus$ gives better $k$-space resolution, which is favorable for measuring the low-$k$ (IR) distribution. We use $\focus = \SI{20}{\hertz}$ (\SI{40}{\hertz}) to study the IR (UV) dynamics. As a compromise, we use $\focus = \SI{30}{\hertz}$ to observe the full $n_k$ distributions shown in the left panel of Fig.~\ref{fig:1}{B}.

\begin{figure}[b!]
\centering\includegraphics[width=1\columnwidth]{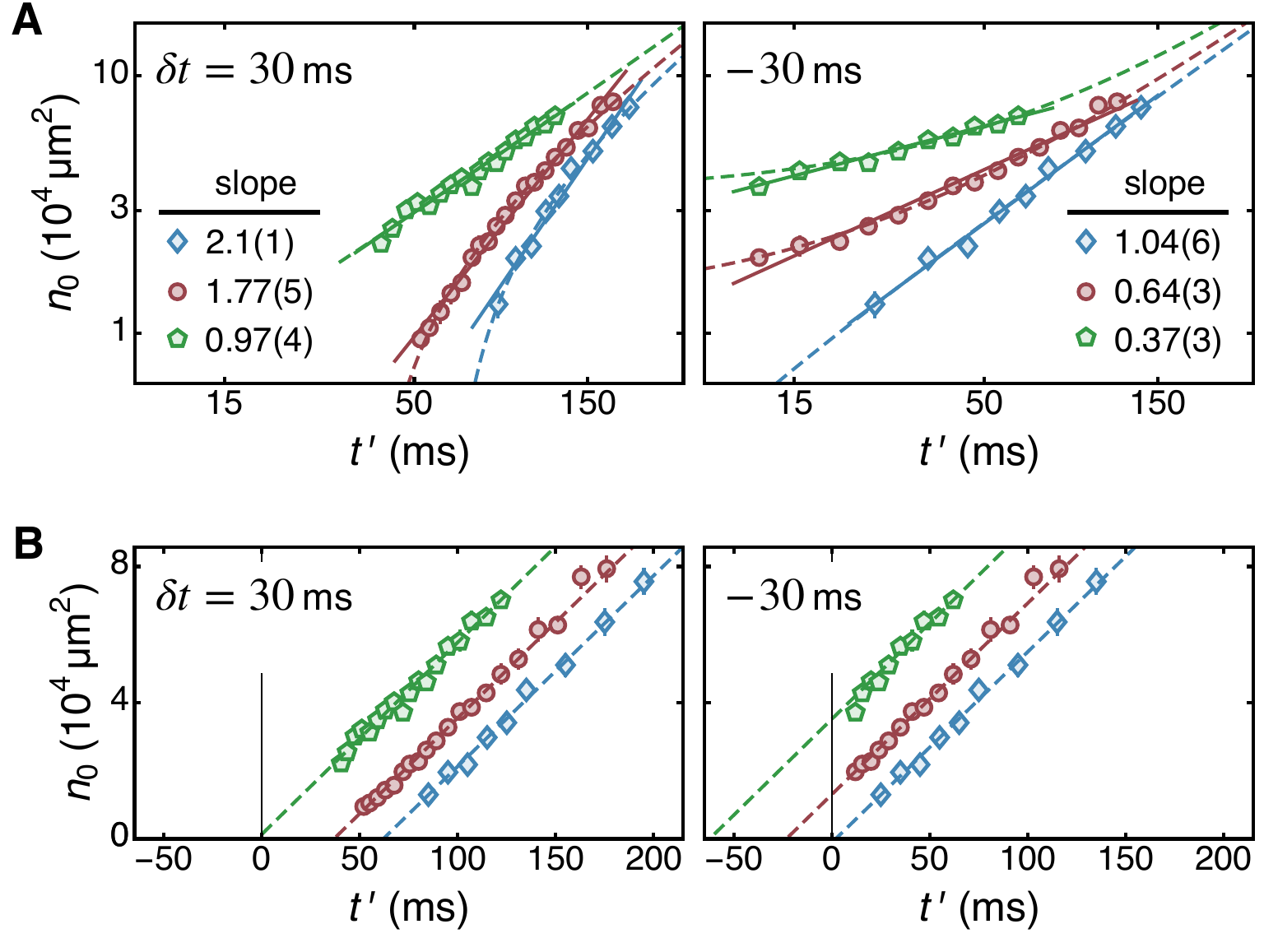}
 \caption{\textbf{Effects of arbitrary shifts of the laboratory time, $t \rightarrow t' = t + \delta t$, illustrated for our IR data.} (\textbf{A}) Shifting by $\delta t = \pm \SI{30}{\ms}$ results in different prescaling exponents, approximated by the slopes of power-law fits (solid lines).  (\textbf{B}) In our analysis method, $\delta t$ is simply absorbed by a shift in $\tstar$, so the correct $\tuni$ and $\alpha$ are recovered. The dashed lines show the same $n_0 \propto \tuni$ curves in (A) and (B).}
 \label{fig:si:1}
\end{figure}

\subsection*{Insensitivity to laboratory clock shifts}

In our experiments $t=0$ is naturally defined by the moment we turn on the interactions and the relaxation starts, while in some other protocols the preparation of a far-from-equilibrium state and the onset of relaxation may not be as unambiguously separated. However, accurate determination of the onset of the (initially non-universal) relaxation is neither sufficient nor necessary for the correct measurement of the universal scaling exponents. Even for the `correct' $t=0$, in general $\tstar \neq 0$ and for non-infinite $t$ one observes prescaling. On the other hand, in our analysis the correct determination of the relevant $\tuni$ and the universal scaling exponents is independent of how $t=0$ is assigned. 

We illustrate this in Fig.~\ref{fig:si:1} for two shifts of the laboratory-clock time, $t$ $\rightarrow t' = t + \delta t$, with arbitrary $\delta t = \pm\SI{30}{\ms}$. These shifts lead to dramatic changes in the prescaling exponents observed in the same data (Fig.~\ref{fig:si:1}{A}), but in our analysis they are simply absorbed by the deduced $\tstar$ values, so that the correct $\tuni$ and the universal scaling exponent are recovered (Fig.~\ref{fig:si:1}{B}).

%%%%%%%%%
\subsection*{Additional panels for Fig.~\ref{fig:3}}

\begin{figure}[b!]
\centering\includegraphics[width=1\columnwidth]{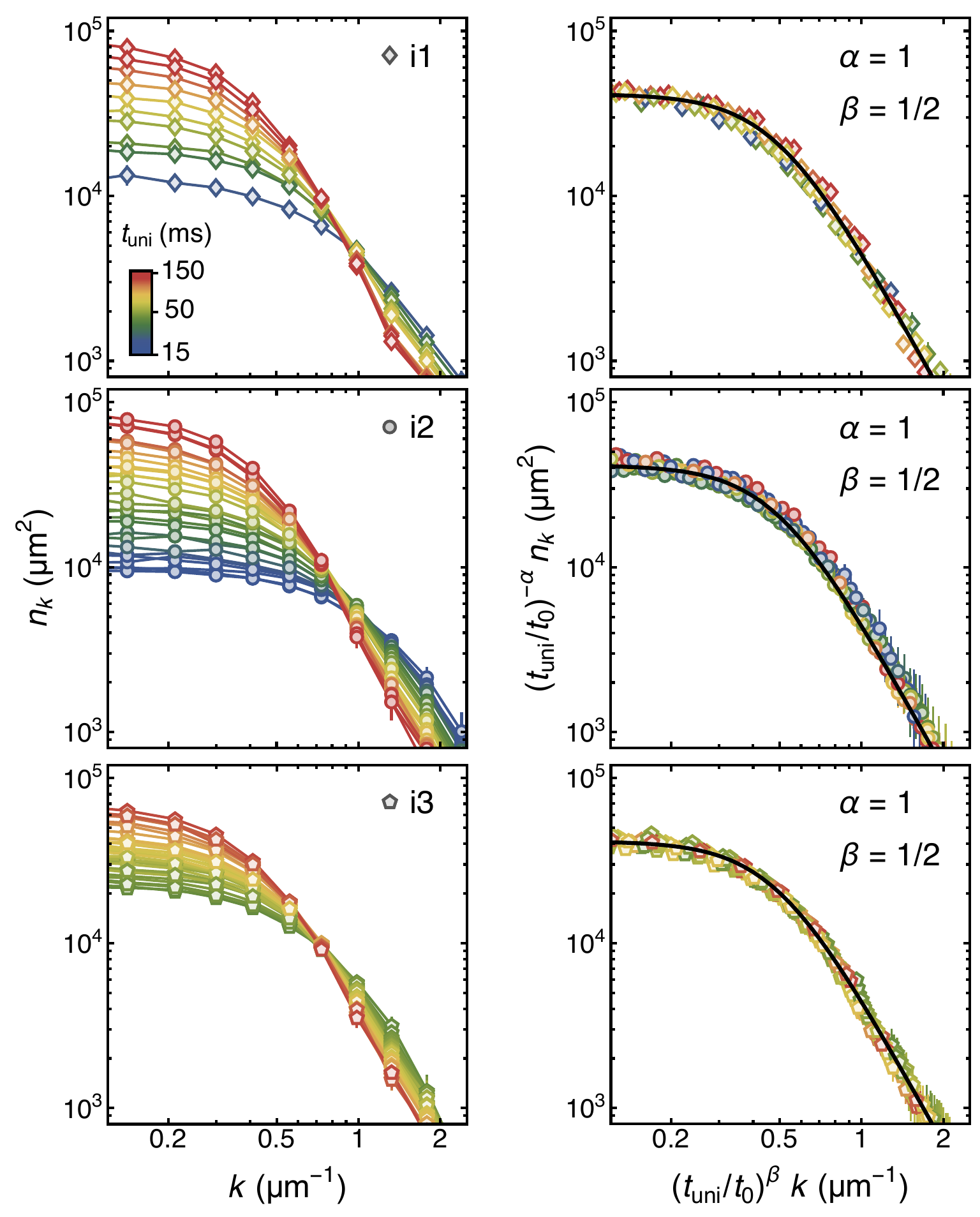}
 \caption{
\textbf{Data from Fig.~\ref{fig:3}B shown separately for the three initial states i1--i3.} In the right panels the curves are collapsed according to Eq.~\eqref{eq:1} with $t\rightarrow\tuni$, $\alpha=1$, and $\beta=1/2$, and the solid lines are the same as the solid line in the right panel of Fig.~\ref{fig:3}B. 
}
 \label{fig:si:2}
\end{figure}

Fig.~\ref{fig:3}{B} shows the simultaneous collapse of the IR $n_k$ curves for the initial states i1--i3, when scaled according to Eq.~\eqref{eq:1} with $t\rightarrow\tuni$, $\alpha=1$, and $\beta=1/2$. In Fig.~\ref{fig:si:2} we separately show the scaling of $n_k$ for each initial state.

\newpage
\subsection*{Additional panels for Fig.~\ref{fig:4}:\texorpdfstring{\\}{} Definition of \texorpdfstring{$k_\varepsilon$}{k-epsilon} and the partial collapse of the UV data} 

\begin{figure}[h!]
\centering
\includegraphics[width=1\columnwidth]{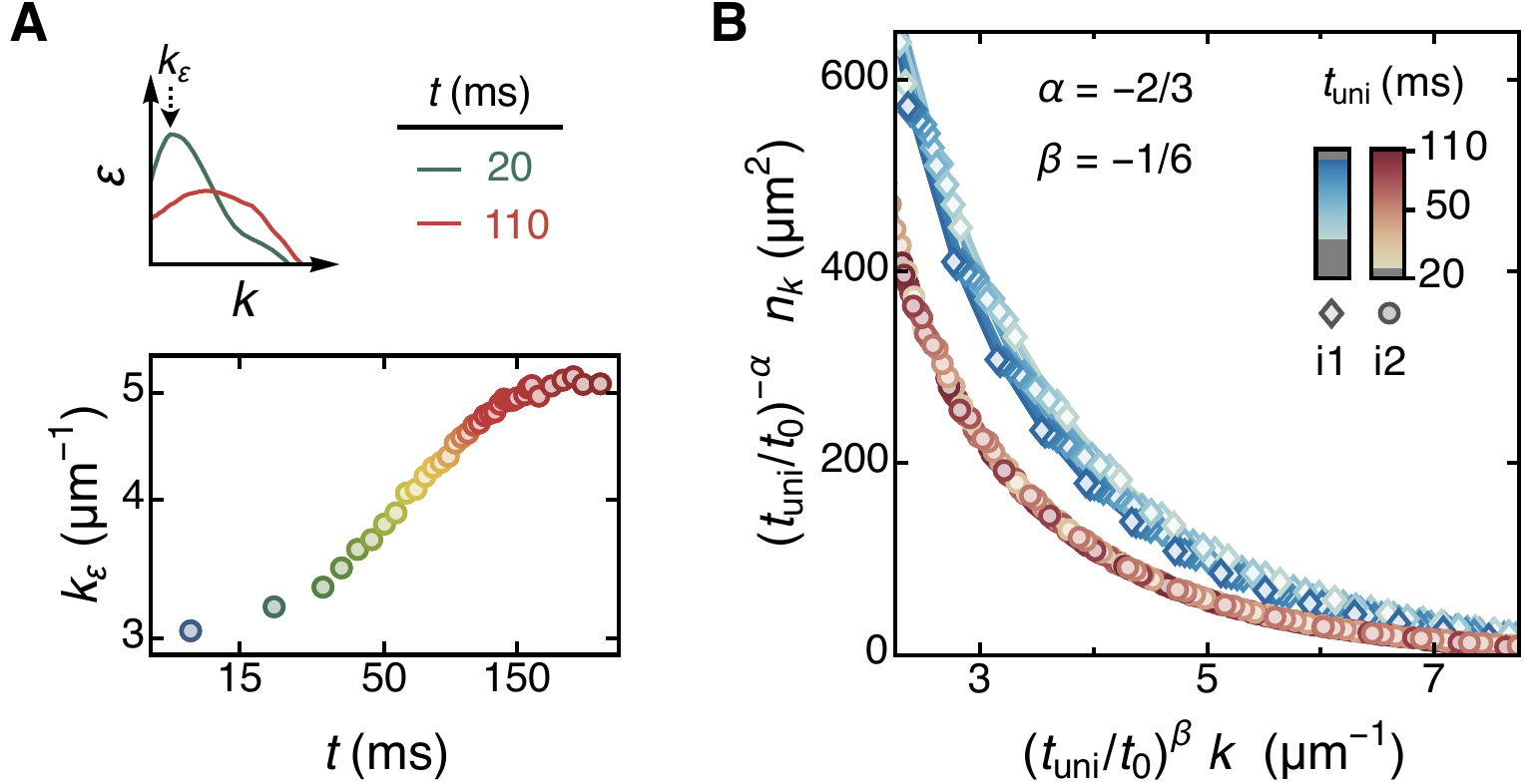}
 \caption{
\textbf{Additional UV panels.} %
(\textbf{A}) The top panel shows two examples of the energy spectrum $\varepsilon(k)$, from the i2 data, and how $k_\varepsilon$ is defined. The bottom panel shows a log-log plot of the full time evolution of $k_\varepsilon$ for i2. 
(\textbf{B}) The UV momentum distributions collapse separately for i1 and i2 when scaled according to Eq.~\eqref{eq:1} with  $t \rightarrow \tuni$, $\alpha = -2/3$, and $\beta = -1/6$; $\tref = \SI{80}{\ms}$ is an arbitrary reference~time. 
}
 \label{fig:si:3}
\end{figure}

In Fig.~\ref{fig:si:3}A we show the definition of $k_\varepsilon$ and its full evolution for i2.

In Fig.~\ref{fig:si:3}B we show a partial collapse of the UV distributions shown in Fig.~\ref{fig:4}{B}. Here, just by setting $t \rightarrow \tuni$, $\alpha=-2/3$, and $\beta=-1/6$, the i1 and i2 data collapse onto two separate curves, which do not have the same normalization; since $\varepsilon(k) \propto n_k$, their integrals are proportional to the total energies $E({\rm i1})$ and $E({\rm i2})$.

\setcounter{figure}{2}

%%%%%%%%%%%%%%%%%%%%%%%%%%%%%%%%%%%%%%%% 

\begin{thebibliography}{54}%
\makeatletter
\providecommand \@ifxundefined [1]{%
 \@ifx{#1\undefined}
}%
\providecommand \@ifnum [1]{%
 \ifnum #1\expandafter \@firstoftwo
 \else \expandafter \@secondoftwo
 \fi
}%
\providecommand \@ifx [1]{%
 \ifx #1\expandafter \@firstoftwo
 \else \expandafter \@secondoftwo
 \fi
}%
\providecommand \natexlab [1]{#1}%
\providecommand \enquote  [1]{``#1''}%
\providecommand \bibnamefont  [1]{#1}%
\providecommand \bibfnamefont [1]{#1}%
\providecommand \citenamefont [1]{#1}%
\providecommand \href@noop [0]{\@secondoftwo}%
\providecommand \href [0]{\begingroup \@sanitize@url \@href}%
\providecommand \@href[1]{\@@startlink{#1}\@@href}%
\providecommand \@@href[1]{\endgroup#1\@@endlink}%
\providecommand \@sanitize@url [0]{\catcode `\\12\catcode `\$12\catcode
  `\&12\catcode `\#12\catcode `\^12\catcode `\_12\catcode `\%12\relax}%
\providecommand \@@startlink[1]{}%
\providecommand \@@endlink[0]{}%
\providecommand \url  [0]{\begingroup\@sanitize@url \@url }%
\providecommand \@url [1]{\endgroup\@href {#1}{\urlprefix }}%
\providecommand \urlprefix  [0]{URL }%
\providecommand \Eprint [0]{\href }%
\providecommand \doibase [0]{https://doi.org/}%
\providecommand \selectlanguage [0]{\@gobble}%
\providecommand \bibinfo  [0]{\@secondoftwo}%
\providecommand \bibfield  [0]{\@secondoftwo}%
\providecommand \translation [1]{[#1]}%
\providecommand \BibitemOpen [0]{}%
\providecommand \bibitemStop [0]{}%
\providecommand \bibitemNoStop [0]{.\EOS\space}%
\providecommand \EOS [0]{\spacefactor3000\relax}%
\providecommand \BibitemShut  [1]{\csname bibitem#1\endcsname}%
\let\auto@bib@innerbib\@empty
%</preamble>
\bibitem [{\citenamefont {Hohenberg}\ and\ \citenamefont
  {Halperin}(1977)}]{Hohenberg:1977}%
  \BibitemOpen
  \bibfield  {author} {\bibinfo {author} {\bibfnamefont {P.~C.}\ \bibnamefont
  {Hohenberg}}\ and\ \bibinfo {author} {\bibfnamefont {B.~I.}\ \bibnamefont
  {Halperin}},\ }\bibfield  {title} {\bibinfo {title} {Theory of dynamic
  critical phenomena},\ }\href {https://doi.org/10.1103/RevModPhys.49.435}
  {\bibfield  {journal} {\bibinfo  {journal} {Rev. Mod. Phys.}\ }\textbf
  {\bibinfo {volume} {49}},\ \bibinfo {pages} {435} (\bibinfo {year}
  {1977})}\BibitemShut {NoStop}%
\bibitem [{\citenamefont {Cross}\ and\ \citenamefont
  {Hohenberg}(1993)}]{Cross:1993}%
  \BibitemOpen
  \bibfield  {author} {\bibinfo {author} {\bibfnamefont {M.~C.}\ \bibnamefont
  {Cross}}\ and\ \bibinfo {author} {\bibfnamefont {P.~C.}\ \bibnamefont
  {Hohenberg}},\ }\bibfield  {title} {\bibinfo {title} {Pattern formation
  outside of equilibrium},\ }\href {https://doi.org/10.1103/RevModPhys.65.851}
  {\bibfield  {journal} {\bibinfo  {journal} {Rev. Mod. Phys.}\ }\textbf
  {\bibinfo {volume} {65}},\ \bibinfo {pages} {851} (\bibinfo {year}
  {1993})}\BibitemShut {NoStop}%
\bibitem [{\citenamefont {Ramaswamy}(2010)}]{Ramaswamy:2010}%
  \BibitemOpen
  \bibfield  {author} {\bibinfo {author} {\bibfnamefont {S.}~\bibnamefont
  {Ramaswamy}},\ }\bibfield  {title} {\bibinfo {title} {{The Mechanics and
  Statistics of Active Matter}},\ }\href
  {https://doi.org/10.1146/annurev-conmatphys-070909-104101} {\bibfield
  {journal} {\bibinfo  {journal} {Annu. Rev. Condens. Matter Phys.}\ }\textbf
  {\bibinfo {volume} {1}},\ \bibinfo {pages} {323} (\bibinfo {year}
  {2010})}\BibitemShut {NoStop}%
\bibitem [{\citenamefont {Bowick}\ \emph {et~al.}(2022)\citenamefont {Bowick},
  \citenamefont {Fakhri}, \citenamefont {Marchetti},\ and\ \citenamefont
  {Ramaswamy}}]{Bowick:2022}%
  \BibitemOpen
  \bibfield  {author} {\bibinfo {author} {\bibfnamefont {M.~J.}\ \bibnamefont
  {Bowick}}, \bibinfo {author} {\bibfnamefont {N.}~\bibnamefont {Fakhri}},
  \bibinfo {author} {\bibfnamefont {M.~C.}\ \bibnamefont {Marchetti}},\ and\
  \bibinfo {author} {\bibfnamefont {S.}~\bibnamefont {Ramaswamy}},\ }\bibfield
  {title} {\bibinfo {title} {{Symmetry, Thermodynamics, and Topology in Active
  Matter}},\ }\href {https://doi.org/10.1103/PhysRevX.12.010501} {\bibfield
  {journal} {\bibinfo  {journal} {Phys. Rev. X}\ }\textbf {\bibinfo {volume}
  {12}},\ \bibinfo {pages} {010501} (\bibinfo {year} {2022})}\BibitemShut
  {NoStop}%
\bibitem [{\citenamefont {{Bray}}(2002)}]{Bray:2002}%
  \BibitemOpen
  \bibfield  {author} {\bibinfo {author} {\bibfnamefont {A.~J.}\ \bibnamefont
  {{Bray}}},\ }\bibfield  {title} {\bibinfo {title} {{Theory of phase-ordering
  kinetics}},\ }\href {https://doi.org/10.1080/00018730110117433} {\bibfield
  {journal} {\bibinfo  {journal} {Adv. Phys.}\ }\textbf {\bibinfo {volume}
  {51}},\ \bibinfo {pages} {481} (\bibinfo {year} {2002})}\BibitemShut
  {NoStop}%
\bibitem [{\citenamefont {Cugliandolo}(2015)}]{Cugliandolo:2015}%
  \BibitemOpen
  \bibfield  {author} {\bibinfo {author} {\bibfnamefont {L.~F.}\ \bibnamefont
  {Cugliandolo}},\ }\bibfield  {title} {\bibinfo {title} {Coarsening
  phenomena},\ }\href
  {https://doi.org/https://doi.org/10.1016/j.crhy.2015.02.005} {\bibfield
  {journal} {\bibinfo  {journal} {C. R. Phys.}\ }\textbf {\bibinfo {volume}
  {16}},\ \bibinfo {pages} {257} (\bibinfo {year} {2015})}\BibitemShut
  {NoStop}%
\bibitem [{\citenamefont {{Damle}}\ \emph {et~al.}(1996)\citenamefont
  {{Damle}}, \citenamefont {{Majumdar}},\ and\ \citenamefont
  {{Sachdev}}}]{Damle:1996}%
  \BibitemOpen
  \bibfield  {author} {\bibinfo {author} {\bibfnamefont {K.}~\bibnamefont
  {{Damle}}}, \bibinfo {author} {\bibfnamefont {S.~N.}\ \bibnamefont
  {{Majumdar}}},\ and\ \bibinfo {author} {\bibfnamefont {S.}~\bibnamefont
  {{Sachdev}}},\ }\bibfield  {title} {\bibinfo {title} {{Phase ordering
  kinetics of the Bose gas}},\ }\href
  {https://doi.org/10.1103/PhysRevA.54.5037} {\bibfield  {journal} {\bibinfo
  {journal} {\pra}\ }\textbf {\bibinfo {volume} {54}},\ \bibinfo {pages} {5037}
  (\bibinfo {year} {1996})}\BibitemShut {NoStop}%
\bibitem [{\citenamefont {Berges}\ \emph {et~al.}(2008)\citenamefont {Berges},
  \citenamefont {Rothkopf},\ and\ \citenamefont {Schmidt}}]{Berges:2008}%
  \BibitemOpen
  \bibfield  {author} {\bibinfo {author} {\bibfnamefont {J.}~\bibnamefont
  {Berges}}, \bibinfo {author} {\bibfnamefont {A.}~\bibnamefont {Rothkopf}},\
  and\ \bibinfo {author} {\bibfnamefont {J.}~\bibnamefont {Schmidt}},\
  }\bibfield  {title} {\bibinfo {title} {{Nonthermal Fixed Points: Effective
  Weak Coupling for Strongly Correlated Systems Far from Equilibrium}},\ }\href
  {https://doi.org/10.1103/PhysRevLett.101.041603} {\bibfield  {journal}
  {\bibinfo  {journal} {Phys. Rev. Lett.}\ }\textbf {\bibinfo {volume} {101}},\
  \bibinfo {pages} {041603} (\bibinfo {year} {2008})}\BibitemShut {NoStop}%
\bibitem [{\citenamefont {Polkovnikov}\ \emph {et~al.}(2011)\citenamefont
  {Polkovnikov}, \citenamefont {Sengupta}, \citenamefont {Silva},\ and\
  \citenamefont {Vengalattore}}]{Polkovnikov:2011}%
  \BibitemOpen
  \bibfield  {author} {\bibinfo {author} {\bibfnamefont {A.}~\bibnamefont
  {Polkovnikov}}, \bibinfo {author} {\bibfnamefont {K.}~\bibnamefont
  {Sengupta}}, \bibinfo {author} {\bibfnamefont {A.}~\bibnamefont {Silva}},\
  and\ \bibinfo {author} {\bibfnamefont {M.}~\bibnamefont {Vengalattore}},\
  }\bibfield  {title} {\bibinfo {title} {\textit{Colloquium}: Nonequilibrium
  dynamics of closed interacting quantum systems},\ }\href
  {https://doi.org/10.1103/RevModPhys.83.863} {\bibfield  {journal} {\bibinfo
  {journal} {Rev. Mod. Phys.}\ }\textbf {\bibinfo {volume} {83}},\ \bibinfo
  {pages} {863} (\bibinfo {year} {2011})}\BibitemShut {NoStop}%
\bibitem [{\citenamefont {{Mikheev}}\ \emph {et~al.}(2023)\citenamefont
  {{Mikheev}}, \citenamefont {{Siovitz}},\ and\ \citenamefont
  {{Gasenzer}}}]{Mikheev:2023}%
  \BibitemOpen
  \bibfield  {author} {\bibinfo {author} {\bibfnamefont {A.~N.}\ \bibnamefont
  {{Mikheev}}}, \bibinfo {author} {\bibfnamefont {I.}~\bibnamefont
  {{Siovitz}}},\ and\ \bibinfo {author} {\bibfnamefont {T.}~\bibnamefont
  {{Gasenzer}}},\ }\bibfield  {title} {\bibinfo {title} {{Universal dynamics
  and non-thermal fixed points in quantum fluids far from equilibrium}},\
  }\href {https://doi.org/10.1140/epjs/s11734-023-00974-7} {\bibfield
  {journal} {\bibinfo  {journal} {Eur. Phys. J. Spec. Top.}\ } (\bibinfo {year}
  {2023})}\BibitemShut {NoStop}%
\bibitem [{\citenamefont {{Berges}}\ and\ \citenamefont
  {{Hoffmeister}}(2009)}]{Berges:2009}%
  \BibitemOpen
  \bibfield  {author} {\bibinfo {author} {\bibfnamefont {J.}~\bibnamefont
  {{Berges}}}\ and\ \bibinfo {author} {\bibfnamefont {G.}~\bibnamefont
  {{Hoffmeister}}},\ }\bibfield  {title} {\bibinfo {title} {{Nonthermal fixed
  points and the functional renormalization group}},\ }\href
  {https://doi.org/10.1016/j.nuclphysb.2008.12.017} {\bibfield  {journal}
  {\bibinfo  {journal} {Nucl. Phys. B}\ }\textbf {\bibinfo {volume} {813}},\
  \bibinfo {pages} {383} (\bibinfo {year} {2009})}\BibitemShut {NoStop}%
\bibitem [{\citenamefont {{Micha}}\ and\ \citenamefont
  {{Tkachev}}(2004)}]{Micha:2004}%
  \BibitemOpen
  \bibfield  {author} {\bibinfo {author} {\bibfnamefont {R.}~\bibnamefont
  {{Micha}}}\ and\ \bibinfo {author} {\bibfnamefont {I.~I.}\ \bibnamefont
  {{Tkachev}}},\ }\bibfield  {title} {\bibinfo {title} {{Turbulent
  thermalization}},\ }\href {https://doi.org/10.1103/PhysRevD.70.043538}
  {\bibfield  {journal} {\bibinfo  {journal} {\prd}\ }\textbf {\bibinfo
  {volume} {70}},\ \bibinfo {eid} {043538} (\bibinfo {year}
  {2004})}\BibitemShut {NoStop}%
\bibitem [{\citenamefont {{Berges}}\ \emph {et~al.}(2014)\citenamefont
  {{Berges}}, \citenamefont {{Boguslavski}}, \citenamefont {{Schlichting}},\
  and\ \citenamefont {{Venugopalan}}}]{Berges:2014b}%
  \BibitemOpen
  \bibfield  {author} {\bibinfo {author} {\bibfnamefont {J.}~\bibnamefont
  {{Berges}}}, \bibinfo {author} {\bibfnamefont {K.}~\bibnamefont
  {{Boguslavski}}}, \bibinfo {author} {\bibfnamefont {S.}~\bibnamefont
  {{Schlichting}}},\ and\ \bibinfo {author} {\bibfnamefont {R.}~\bibnamefont
  {{Venugopalan}}},\ }\bibfield  {title} {\bibinfo {title} {{Turbulent
  thermalization process in heavy-ion collisions at ultrarelativistic
  energies}},\ }\href {https://doi.org/10.1103/PhysRevD.89.074011} {\bibfield
  {journal} {\bibinfo  {journal} {\prd}\ }\textbf {\bibinfo {volume} {89}},\
  \bibinfo {eid} {074011} (\bibinfo {year} {2014})}\BibitemShut {NoStop}%
\bibitem [{\citenamefont {Bhattacharyya}\ \emph {et~al.}(2020)\citenamefont
  {Bhattacharyya}, \citenamefont {Rodriguez-Nieva},\ and\ \citenamefont
  {Demler}}]{Bhattacharyya:2020}%
  \BibitemOpen
  \bibfield  {author} {\bibinfo {author} {\bibfnamefont {S.}~\bibnamefont
  {Bhattacharyya}}, \bibinfo {author} {\bibfnamefont {J.~F.}\ \bibnamefont
  {Rodriguez-Nieva}},\ and\ \bibinfo {author} {\bibfnamefont {E.}~\bibnamefont
  {Demler}},\ }\bibfield  {title} {\bibinfo {title} {{Universal Prethermal
  Dynamics in Heisenberg Ferromagnets}},\ }\href
  {https://doi.org/10.1103/PhysRevLett.125.230601} {\bibfield  {journal}
  {\bibinfo  {journal} {Phys. Rev. Lett.}\ }\textbf {\bibinfo {volume} {125}},\
  \bibinfo {pages} {230601} (\bibinfo {year} {2020})}\BibitemShut {NoStop}%
\bibitem [{\citenamefont {{Nowak}}\ \emph {et~al.}(2012)\citenamefont
  {{Nowak}}, \citenamefont {{Schole}}, \citenamefont {{Sexty}},\ and\
  \citenamefont {{Gasenzer}}}]{Nowak:2012}%
  \BibitemOpen
  \bibfield  {author} {\bibinfo {author} {\bibfnamefont {B.}~\bibnamefont
  {{Nowak}}}, \bibinfo {author} {\bibfnamefont {J.}~\bibnamefont {{Schole}}},
  \bibinfo {author} {\bibfnamefont {D.}~\bibnamefont {{Sexty}}},\ and\ \bibinfo
  {author} {\bibfnamefont {T.}~\bibnamefont {{Gasenzer}}},\ }\bibfield  {title}
  {\bibinfo {title} {{Nonthermal fixed points, vortex statistics, and
  superfluid turbulence in an ultracold Bose gas}},\ }\href
  {https://doi.org/10.1103/PhysRevA.85.043627} {\bibfield  {journal} {\bibinfo
  {journal} {\pra}\ }\textbf {\bibinfo {volume} {85}},\ \bibinfo {eid} {043627}
  (\bibinfo {year} {2012})}\BibitemShut {NoStop}%
\bibitem [{\citenamefont {Pi\~neiro Orioli}\ \emph {et~al.}(2015)\citenamefont
  {Pi\~neiro Orioli}, \citenamefont {Boguslavski},\ and\ \citenamefont
  {Berges}}]{PineiroOrioli:2015}%
  \BibitemOpen
  \bibfield  {author} {\bibinfo {author} {\bibfnamefont {A.}~\bibnamefont
  {Pi\~neiro Orioli}}, \bibinfo {author} {\bibfnamefont {K.}~\bibnamefont
  {Boguslavski}},\ and\ \bibinfo {author} {\bibfnamefont {J.}~\bibnamefont
  {Berges}},\ }\bibfield  {title} {\bibinfo {title} {Universal self-similar
  dynamics of relativistic and nonrelativistic field theories near nonthermal
  fixed points},\ }\href {https://doi.org/10.1103/PhysRevD.92.025041}
  {\bibfield  {journal} {\bibinfo  {journal} {Phys. Rev. D}\ }\textbf {\bibinfo
  {volume} {92}},\ \bibinfo {pages} {025041} (\bibinfo {year}
  {2015})}\BibitemShut {NoStop}%
\bibitem [{\citenamefont {Berges}\ \emph {et~al.}(2015)\citenamefont {Berges},
  \citenamefont {Boguslavski}, \citenamefont {Schlichting},\ and\ \citenamefont
  {Venugopalan}}]{Berges:2015}%
  \BibitemOpen
  \bibfield  {author} {\bibinfo {author} {\bibfnamefont {J.}~\bibnamefont
  {Berges}}, \bibinfo {author} {\bibfnamefont {K.}~\bibnamefont {Boguslavski}},
  \bibinfo {author} {\bibfnamefont {S.}~\bibnamefont {Schlichting}},\ and\
  \bibinfo {author} {\bibfnamefont {R.}~\bibnamefont {Venugopalan}},\
  }\bibfield  {title} {\bibinfo {title} {{Universality Far from Equilibrium:
  From Superfluid Bose Gases to Heavy-Ion Collisions}},\ }\href
  {https://doi.org/10.1103/PhysRevLett.114.061601} {\bibfield  {journal}
  {\bibinfo  {journal} {Phys. Rev. Lett.}\ }\textbf {\bibinfo {volume} {114}},\
  \bibinfo {pages} {061601} (\bibinfo {year} {2015})}\BibitemShut {NoStop}%
\bibitem [{\citenamefont {{Karl}}\ and\ \citenamefont
  {{Gasenzer}}(2017)}]{Karl:2017}%
  \BibitemOpen
  \bibfield  {author} {\bibinfo {author} {\bibfnamefont {M.}~\bibnamefont
  {{Karl}}}\ and\ \bibinfo {author} {\bibfnamefont {T.}~\bibnamefont
  {{Gasenzer}}},\ }\bibfield  {title} {\bibinfo {title} {{Strongly anomalous
  non-thermal fixed point in a quenched two-dimensional Bose gas}},\ }\href
  {https://doi.org/10.1088/1367-2630/aa7eeb} {\bibfield  {journal} {\bibinfo
  {journal} {New J. Phys.}\ }\textbf {\bibinfo {volume} {19}},\ \bibinfo {eid}
  {093014} (\bibinfo {year} {2017})}\BibitemShut {NoStop}%
\bibitem [{\citenamefont {Chantesana}\ \emph {et~al.}(2019)\citenamefont
  {Chantesana}, \citenamefont {Pi\~neiro Orioli},\ and\ \citenamefont
  {Gasenzer}}]{Chantesana:2019}%
  \BibitemOpen
  \bibfield  {author} {\bibinfo {author} {\bibfnamefont {I.}~\bibnamefont
  {Chantesana}}, \bibinfo {author} {\bibfnamefont {A.}~\bibnamefont {Pi\~neiro
  Orioli}},\ and\ \bibinfo {author} {\bibfnamefont {T.}~\bibnamefont
  {Gasenzer}},\ }\bibfield  {title} {\bibinfo {title} {{Kinetic theory of
  nonthermal fixed points in a Bose gas}},\ }\href
  {https://doi.org/10.1103/PhysRevA.99.043620} {\bibfield  {journal} {\bibinfo
  {journal} {Phys. Rev. A}\ }\textbf {\bibinfo {volume} {99}},\ \bibinfo
  {pages} {043620} (\bibinfo {year} {2019})}\BibitemShut {NoStop}%
\bibitem [{\citenamefont {{Chatrchyan}}\ \emph {et~al.}(2021)\citenamefont
  {{Chatrchyan}}, \citenamefont {{Geier}}, \citenamefont {{Oberthaler}},
  \citenamefont {{Berges}},\ and\ \citenamefont {{Hauke}}}]{Chatrchyan:2021}%
  \BibitemOpen
  \bibfield  {author} {\bibinfo {author} {\bibfnamefont {A.}~\bibnamefont
  {{Chatrchyan}}}, \bibinfo {author} {\bibfnamefont {K.~T.}\ \bibnamefont
  {{Geier}}}, \bibinfo {author} {\bibfnamefont {M.~K.}\ \bibnamefont
  {{Oberthaler}}}, \bibinfo {author} {\bibfnamefont {J.}~\bibnamefont
  {{Berges}}},\ and\ \bibinfo {author} {\bibfnamefont {P.}~\bibnamefont
  {{Hauke}}},\ }\bibfield  {title} {\bibinfo {title} {{Analog cosmological
  reheating in an ultracold Bose gas}},\ }\href
  {https://doi.org/10.1103/PhysRevA.104.023302} {\bibfield  {journal} {\bibinfo
   {journal} {\pra}\ }\textbf {\bibinfo {volume} {104}},\ \bibinfo {eid}
  {023302} (\bibinfo {year} {2021})}\BibitemShut {NoStop}%
\bibitem [{\citenamefont {Groszek}\ \emph {et~al.}(2021)\citenamefont
  {Groszek}, \citenamefont {Comaron}, \citenamefont {Proukakis},\ and\
  \citenamefont {Billam}}]{Groszek:2021}%
  \BibitemOpen
  \bibfield  {author} {\bibinfo {author} {\bibfnamefont {A.~J.}\ \bibnamefont
  {Groszek}}, \bibinfo {author} {\bibfnamefont {P.}~\bibnamefont {Comaron}},
  \bibinfo {author} {\bibfnamefont {N.~P.}\ \bibnamefont {Proukakis}},\ and\
  \bibinfo {author} {\bibfnamefont {T.~P.}\ \bibnamefont {Billam}},\ }\bibfield
   {title} {\bibinfo {title} {{Crossover in the dynamical critical exponent of
  a quenched two-dimensional Bose gas}},\ }\href
  {https://doi.org/10.1103/PhysRevResearch.3.013212} {\bibfield  {journal}
  {\bibinfo  {journal} {Phys. Rev. Res.}\ }\textbf {\bibinfo {volume} {3}},\
  \bibinfo {pages} {013212} (\bibinfo {year} {2021})}\BibitemShut {NoStop}%
\bibitem [{\citenamefont {{Gresista}}\ \emph {et~al.}(2022)\citenamefont
  {{Gresista}}, \citenamefont {{Zache}},\ and\ \citenamefont
  {{Berges}}}]{Gresista:2022}%
  \BibitemOpen
  \bibfield  {author} {\bibinfo {author} {\bibfnamefont {L.}~\bibnamefont
  {{Gresista}}}, \bibinfo {author} {\bibfnamefont {T.~V.}\ \bibnamefont
  {{Zache}}},\ and\ \bibinfo {author} {\bibfnamefont {J.}~\bibnamefont
  {{Berges}}},\ }\bibfield  {title} {\bibinfo {title} {{Dimensional crossover
  for universal scaling far from equilibrium}},\ }\href
  {https://doi.org/10.1103/PhysRevA.105.013320} {\bibfield  {journal} {\bibinfo
   {journal} {\pra}\ }\textbf {\bibinfo {volume} {105}},\ \bibinfo {eid}
  {013320} (\bibinfo {year} {2022})}\BibitemShut {NoStop}%
\bibitem [{\citenamefont {Chaikin}\ and\ \citenamefont
  {Lubensky}(1995)}]{Chaikin:1995}%
  \BibitemOpen
  \bibfield  {author} {\bibinfo {author} {\bibfnamefont {P.~M.}\ \bibnamefont
  {Chaikin}}\ and\ \bibinfo {author} {\bibfnamefont {T.~C.}\ \bibnamefont
  {Lubensky}},\ }\href {https://doi.org/10.1017/CBO9780511813467} {\emph
  {\bibinfo {title} {Principles of Condensed Matter Physics}}}\ (\bibinfo
  {publisher} {Cambridge University Press},\ \bibinfo {year}
  {1995})\BibitemShut {NoStop}%
\bibitem [{\citenamefont {{Pr{\"u}fer}}\ \emph {et~al.}(2018)\citenamefont
  {{Pr{\"u}fer}}, \citenamefont {{Kunkel}}, \citenamefont {{Strobel}},
  \citenamefont {{Lannig}}, \citenamefont {{Linnemann}}, \citenamefont
  {{Schmied}}, \citenamefont {{Berges}}, \citenamefont {{Gasenzer}},\ and\
  \citenamefont {{Oberthaler}}}]{Prufer:2018}%
  \BibitemOpen
  \bibfield  {author} {\bibinfo {author} {\bibfnamefont {M.}~\bibnamefont
  {{Pr{\"u}fer}}}, \bibinfo {author} {\bibfnamefont {P.}~\bibnamefont
  {{Kunkel}}}, \bibinfo {author} {\bibfnamefont {H.}~\bibnamefont {{Strobel}}},
  \bibinfo {author} {\bibfnamefont {S.}~\bibnamefont {{Lannig}}}, \bibinfo
  {author} {\bibfnamefont {D.}~\bibnamefont {{Linnemann}}}, \bibinfo {author}
  {\bibfnamefont {C.-M.}\ \bibnamefont {{Schmied}}}, \bibinfo {author}
  {\bibfnamefont {J.}~\bibnamefont {{Berges}}}, \bibinfo {author}
  {\bibfnamefont {T.}~\bibnamefont {{Gasenzer}}},\ and\ \bibinfo {author}
  {\bibfnamefont {M.~K.}\ \bibnamefont {{Oberthaler}}},\ }\bibfield  {title}
  {\bibinfo {title} {{Observation of universal dynamics in a spinor Bose gas
  far from equilibrium}},\ }\href {https://doi.org/10.1038/s41586-018-0659-0}
  {\bibfield  {journal} {\bibinfo  {journal} {Nature}\ }\textbf {\bibinfo
  {volume} {563}},\ \bibinfo {pages} {217} (\bibinfo {year}
  {2018})}\BibitemShut {NoStop}%
\bibitem [{\citenamefont {Erne}\ \emph {et~al.}(2018)\citenamefont {Erne},
  \citenamefont {B{\"u}cker}, \citenamefont {Gasenzer}, \citenamefont
  {Berges},\ and\ \citenamefont {Schmiedmayer}}]{Erne:2018}%
  \BibitemOpen
  \bibfield  {author} {\bibinfo {author} {\bibfnamefont {S.}~\bibnamefont
  {Erne}}, \bibinfo {author} {\bibfnamefont {R.}~\bibnamefont {B{\"u}cker}},
  \bibinfo {author} {\bibfnamefont {T.}~\bibnamefont {Gasenzer}}, \bibinfo
  {author} {\bibfnamefont {J.}~\bibnamefont {Berges}},\ and\ \bibinfo {author}
  {\bibfnamefont {J.}~\bibnamefont {Schmiedmayer}},\ }\bibfield  {title}
  {\bibinfo {title} {Universal dynamics in an isolated one-dimensional {B}ose
  gas far from equilibrium},\ }\href
  {https://doi.org/10.1038/s41586-018-0667-0} {\bibfield  {journal} {\bibinfo
  {journal} {Nature}\ }\textbf {\bibinfo {volume} {563}},\ \bibinfo {pages}
  {225} (\bibinfo {year} {2018})}\BibitemShut {NoStop}%
\bibitem [{\citenamefont {Glidden}\ \emph {et~al.}(2021)\citenamefont
  {Glidden}, \citenamefont {Eigen}, \citenamefont {Dogra}, \citenamefont
  {Hilker}, \citenamefont {Smith},\ and\ \citenamefont
  {Hadzibabic}}]{Glidden:2021}%
  \BibitemOpen
  \bibfield  {author} {\bibinfo {author} {\bibfnamefont {J.~A.~P.}\
  \bibnamefont {Glidden}}, \bibinfo {author} {\bibfnamefont {C.}~\bibnamefont
  {Eigen}}, \bibinfo {author} {\bibfnamefont {L.~H.}\ \bibnamefont {Dogra}},
  \bibinfo {author} {\bibfnamefont {T.~A.}\ \bibnamefont {Hilker}}, \bibinfo
  {author} {\bibfnamefont {R.~P.}\ \bibnamefont {Smith}},\ and\ \bibinfo
  {author} {\bibfnamefont {Z.}~\bibnamefont {Hadzibabic}},\ }\bibfield  {title}
  {\bibinfo {title} {{Bidirectional dynamic scaling in an isolated Bose gas far
  from equilibrium}},\ }\href {https://doi.org/10.1038/s41567-020-01114-x}
  {\bibfield  {journal} {\bibinfo  {journal} {Nat. Phys.}\ }\textbf {\bibinfo
  {volume} {17}},\ \bibinfo {pages} {457} (\bibinfo {year} {2021})}\BibitemShut
  {NoStop}%
\bibitem [{\citenamefont {{Garc{\'\i}a-Orozco}}\ \emph
  {et~al.}(2022)\citenamefont {{Garc{\'\i}a-Orozco}}, \citenamefont
  {{Madeira}}, \citenamefont {{Moreno-Armijos}}, \citenamefont {{Fritsch}},
  \citenamefont {{Tavares}}, \citenamefont {{Castilho}}, \citenamefont
  {{Cidrim}}, \citenamefont {{Roati}},\ and\ \citenamefont
  {{Bagnato}}}]{Orozco:2022}%
  \BibitemOpen
  \bibfield  {author} {\bibinfo {author} {\bibfnamefont {A.~D.}\ \bibnamefont
  {{Garc{\'\i}a-Orozco}}}, \bibinfo {author} {\bibfnamefont {L.}~\bibnamefont
  {{Madeira}}}, \bibinfo {author} {\bibfnamefont {M.~A.}\ \bibnamefont
  {{Moreno-Armijos}}}, \bibinfo {author} {\bibfnamefont {A.~R.}\ \bibnamefont
  {{Fritsch}}}, \bibinfo {author} {\bibfnamefont {P.~E.~S.}\ \bibnamefont
  {{Tavares}}}, \bibinfo {author} {\bibfnamefont {P.~C.~M.}\ \bibnamefont
  {{Castilho}}}, \bibinfo {author} {\bibfnamefont {A.}~\bibnamefont
  {{Cidrim}}}, \bibinfo {author} {\bibfnamefont {G.}~\bibnamefont {{Roati}}},\
  and\ \bibinfo {author} {\bibfnamefont {V.~S.}\ \bibnamefont {{Bagnato}}},\
  }\bibfield  {title} {\bibinfo {title} {{Universal dynamics of a turbulent
  superfluid Bose gas}},\ }\href {https://doi.org/10.1103/PhysRevA.106.023314}
  {\bibfield  {journal} {\bibinfo  {journal} {\pra}\ }\textbf {\bibinfo
  {volume} {106}},\ \bibinfo {eid} {023314} (\bibinfo {year}
  {2022})}\BibitemShut {NoStop}%
\bibitem [{\citenamefont {{Lannig}}\ \emph {et~al.}(2023)\citenamefont
  {{Lannig}}, \citenamefont {{Pr{\"u}fer}}, \citenamefont {{Deller}},
  \citenamefont {{Siovitz}}, \citenamefont {{Dreher}}, \citenamefont
  {{Gasenzer}}, \citenamefont {{Strobel}},\ and\ \citenamefont
  {{Oberthaler}}}]{Lannig:2023}%
  \BibitemOpen
  \bibfield  {author} {\bibinfo {author} {\bibfnamefont {S.}~\bibnamefont
  {{Lannig}}}, \bibinfo {author} {\bibfnamefont {M.}~\bibnamefont
  {{Pr{\"u}fer}}}, \bibinfo {author} {\bibfnamefont {Y.}~\bibnamefont
  {{Deller}}}, \bibinfo {author} {\bibfnamefont {I.}~\bibnamefont {{Siovitz}}},
  \bibinfo {author} {\bibfnamefont {J.}~\bibnamefont {{Dreher}}}, \bibinfo
  {author} {\bibfnamefont {T.}~\bibnamefont {{Gasenzer}}}, \bibinfo {author}
  {\bibfnamefont {H.}~\bibnamefont {{Strobel}}},\ and\ \bibinfo {author}
  {\bibfnamefont {M.~K.}\ \bibnamefont {{Oberthaler}}},\ }\bibfield  {title}
  {\bibinfo {title} {{Observation of two non-thermal fixed points for the same
  microscopic symmetry}},\ }\href {https://doi.org/10.48550/arXiv.2306.16497}
  {\bibfield  {journal} {\bibinfo  {journal} {arXiv:2306.16497}\ } (\bibinfo
  {year} {2023})}\BibitemShut {NoStop}%
\bibitem [{\citenamefont {Huh}\ \emph {et~al.}(2024)\citenamefont {Huh},
  \citenamefont {Mukherjee}, \citenamefont {Kwon}, \citenamefont {Seo},
  \citenamefont {Hur}, \citenamefont {Mistakidis}, \citenamefont {Sadeghpour},\
  and\ \citenamefont {Choi}}]{Huh:2024}%
  \BibitemOpen
  \bibfield  {author} {\bibinfo {author} {\bibfnamefont {S.}~\bibnamefont
  {Huh}}, \bibinfo {author} {\bibfnamefont {K.}~\bibnamefont {Mukherjee}},
  \bibinfo {author} {\bibfnamefont {K.}~\bibnamefont {Kwon}}, \bibinfo {author}
  {\bibfnamefont {J.}~\bibnamefont {Seo}}, \bibinfo {author} {\bibfnamefont
  {J.}~\bibnamefont {Hur}}, \bibinfo {author} {\bibfnamefont {S.~I.}\
  \bibnamefont {Mistakidis}}, \bibinfo {author} {\bibfnamefont {H.~R.}\
  \bibnamefont {Sadeghpour}},\ and\ \bibinfo {author} {\bibfnamefont {J.-Y.}\
  \bibnamefont {Choi}},\ }\bibfield  {title} {\bibinfo {title} {{Universality
  class of a spinor Bose--Einstein condensate far from equilibrium}},\ }\href
  {https://doi.org/10.1038/s41567-023-02339-2} {\bibfield  {journal} {\bibinfo
  {journal} {Nat. Phys.}\ } (\bibinfo {year} {2024})}\BibitemShut {NoStop}%
\bibitem [{\citenamefont {{Kendon}}\ \emph {et~al.}(2001)\citenamefont
  {{Kendon}}, \citenamefont {{Cates}}, \citenamefont {{Pagonabarraga}},
  \citenamefont {{Desplat}},\ and\ \citenamefont {{Bladon}}}]{Kendon:2001}%
  \BibitemOpen
  \bibfield  {author} {\bibinfo {author} {\bibfnamefont {V.~M.}\ \bibnamefont
  {{Kendon}}}, \bibinfo {author} {\bibfnamefont {M.~E.}\ \bibnamefont
  {{Cates}}}, \bibinfo {author} {\bibfnamefont {I.}~\bibnamefont
  {{Pagonabarraga}}}, \bibinfo {author} {\bibfnamefont {J.~C.}\ \bibnamefont
  {{Desplat}}},\ and\ \bibinfo {author} {\bibfnamefont {P.}~\bibnamefont
  {{Bladon}}},\ }\bibfield  {title} {\bibinfo {title} {{Inertial effects in
  three-dimensional spinodal decomposition of a symmetric binary fluid mixture:
  a lattice Boltzmann study}},\ }\href
  {https://doi.org/10.1017/S0022112001004682} {\bibfield  {journal} {\bibinfo
  {journal} {J. Fluid Mech.}\ }\textbf {\bibinfo {volume} {440}},\ \bibinfo
  {pages} {147} (\bibinfo {year} {2001})}\BibitemShut {NoStop}%
\bibitem [{\citenamefont {{Mazeliauskas}}\ and\ \citenamefont
  {{Berges}}(2019)}]{Mazeliauskas:2019}%
  \BibitemOpen
  \bibfield  {author} {\bibinfo {author} {\bibfnamefont {A.}~\bibnamefont
  {{Mazeliauskas}}}\ and\ \bibinfo {author} {\bibfnamefont {J.}~\bibnamefont
  {{Berges}}},\ }\bibfield  {title} {\bibinfo {title} {{Prescaling and
  Far-from-Equilibrium Hydrodynamics in the Quark-Gluon Plasma}},\ }\href
  {https://doi.org/10.1103/PhysRevLett.122.122301} {\bibfield  {journal}
  {\bibinfo  {journal} {\prl}\ }\textbf {\bibinfo {volume} {122}},\ \bibinfo
  {eid} {122301} (\bibinfo {year} {2019})}\BibitemShut {NoStop}%
\bibitem [{\citenamefont {Schmied}\ \emph {et~al.}(2019)\citenamefont
  {Schmied}, \citenamefont {Mikheev},\ and\ \citenamefont
  {Gasenzer}}]{Schmied:2019}%
  \BibitemOpen
  \bibfield  {author} {\bibinfo {author} {\bibfnamefont {C.-M.}\ \bibnamefont
  {Schmied}}, \bibinfo {author} {\bibfnamefont {A.~N.}\ \bibnamefont
  {Mikheev}},\ and\ \bibinfo {author} {\bibfnamefont {T.}~\bibnamefont
  {Gasenzer}},\ }\bibfield  {title} {\bibinfo {title} {{Prescaling in a
  Far-from-Equilibrium Bose Gas}},\ }\href
  {https://doi.org/10.1103/PhysRevLett.122.170404} {\bibfield  {journal}
  {\bibinfo  {journal} {Phys. Rev. Lett.}\ }\textbf {\bibinfo {volume} {122}},\
  \bibinfo {pages} {170404} (\bibinfo {year} {2019})}\BibitemShut {NoStop}%
\bibitem [{\citenamefont {{Heller}}\ \emph {et~al.}(2023)\citenamefont
  {{Heller}}, \citenamefont {{Mazeliauskas}},\ and\ \citenamefont
  {{Preis}}}]{Heller:2023}%
  \BibitemOpen
  \bibfield  {author} {\bibinfo {author} {\bibfnamefont {M.~P.}\ \bibnamefont
  {{Heller}}}, \bibinfo {author} {\bibfnamefont {A.}~\bibnamefont
  {{Mazeliauskas}}},\ and\ \bibinfo {author} {\bibfnamefont {T.}~\bibnamefont
  {{Preis}}},\ }\bibfield  {title} {\bibinfo {title} {{Prescaling relaxation to
  nonthermal attractors}},\ }\href {https://doi.org/10.48550/arXiv.2307.07545}
  {\bibfield  {journal} {\bibinfo  {journal} {arXiv:2307.07545}\ } (\bibinfo
  {year} {2023})}\BibitemShut {NoStop}%
\bibitem [{\citenamefont {Chomaz}\ \emph {et~al.}(2015)\citenamefont {Chomaz},
  \citenamefont {Corman}, \citenamefont {Bienaim{\'e}}, \citenamefont
  {Desbuquois}, \citenamefont {Weitenberg}, \citenamefont {Nascimb{\`e}ne},
  \citenamefont {Beugnon},\ and\ \citenamefont {Dalibard}}]{Chomaz:2015}%
  \BibitemOpen
  \bibfield  {author} {\bibinfo {author} {\bibfnamefont {L.}~\bibnamefont
  {Chomaz}}, \bibinfo {author} {\bibfnamefont {L.}~\bibnamefont {Corman}},
  \bibinfo {author} {\bibfnamefont {T.}~\bibnamefont {Bienaim{\'e}}}, \bibinfo
  {author} {\bibfnamefont {R.}~\bibnamefont {Desbuquois}}, \bibinfo {author}
  {\bibfnamefont {C.}~\bibnamefont {Weitenberg}}, \bibinfo {author}
  {\bibfnamefont {S.}~\bibnamefont {Nascimb{\`e}ne}}, \bibinfo {author}
  {\bibfnamefont {J.}~\bibnamefont {Beugnon}},\ and\ \bibinfo {author}
  {\bibfnamefont {J.}~\bibnamefont {Dalibard}},\ }\bibfield  {title} {\bibinfo
  {title} {{Emergence of coherence via transverse condensation in a uniform
  quasi-two-dimensional Bose gas}},\ }\href
  {https://doi.org/10.1038/ncomms7162} {\bibfield  {journal} {\bibinfo
  {journal} {Nat. Commun.}\ }\textbf {\bibinfo {volume} {6}},\ \bibinfo {pages}
  {6162} (\bibinfo {year} {2015})}\BibitemShut {NoStop}%
\bibitem [{\citenamefont {Navon}\ \emph {et~al.}(2021)\citenamefont {Navon},
  \citenamefont {Smith},\ and\ \citenamefont {Hadzibabic}}]{Navon:2021}%
  \BibitemOpen
  \bibfield  {author} {\bibinfo {author} {\bibfnamefont {N.}~\bibnamefont
  {Navon}}, \bibinfo {author} {\bibfnamefont {R.~P.}\ \bibnamefont {Smith}},\
  and\ \bibinfo {author} {\bibfnamefont {Z.}~\bibnamefont {Hadzibabic}},\
  }\bibfield  {title} {\bibinfo {title} {Quantum gases in optical boxes},\
  }\href {https://doi.org/10.1038/s41567-021-01403-z} {\bibfield  {journal}
  {\bibinfo  {journal} {Nat. Phys.}\ }\textbf {\bibinfo {volume} {17}},\
  \bibinfo {pages} {1334} (\bibinfo {year} {2021})}\BibitemShut {NoStop}%
\bibitem [{\citenamefont {{Tung}}\ \emph {et~al.}(2010)\citenamefont {{Tung}},
  \citenamefont {{Lamporesi}}, \citenamefont {{Lobser}}, \citenamefont
  {{Xia}},\ and\ \citenamefont {{Cornell}}}]{Tung:2010}%
  \BibitemOpen
  \bibfield  {author} {\bibinfo {author} {\bibfnamefont {S.}~\bibnamefont
  {{Tung}}}, \bibinfo {author} {\bibfnamefont {G.}~\bibnamefont {{Lamporesi}}},
  \bibinfo {author} {\bibfnamefont {D.}~\bibnamefont {{Lobser}}}, \bibinfo
  {author} {\bibfnamefont {L.}~\bibnamefont {{Xia}}},\ and\ \bibinfo {author}
  {\bibfnamefont {E.~A.}\ \bibnamefont {{Cornell}}},\ }\bibfield  {title}
  {\bibinfo {title} {{Observation of the Presuperfluid Regime in a
  Two-Dimensional Bose Gas}},\ }\href
  {https://doi.org/10.1103/PhysRevLett.105.230408} {\bibfield  {journal}
  {\bibinfo  {journal} {\prl}\ }\textbf {\bibinfo {volume} {105}},\ \bibinfo
  {eid} {230408} (\bibinfo {year} {2010})}\BibitemShut {NoStop}%
\bibitem [{SI()}]{SI}%
  \BibitemOpen
  \bibinfo {note} {{S}ee {S}upplementary {M}aterials.}\BibitemShut {Stop}%
\bibitem [{\citenamefont {Proukakis}(2024)}]{Proukakis:2023}%
  \BibitemOpen
  \bibfield  {author} {\bibinfo {author} {\bibfnamefont {N.~P.}\ \bibnamefont
  {Proukakis}},\ }\bibfield  {title} {\bibinfo {title} {{Universality of
  Bose–Einstein condensation and quenched formation dynamics}},\ }in\ \href
  {https://doi.org/https://doi.org/10.1016/B978-0-323-90800-9.00253-5} {\emph
  {\bibinfo {booktitle} {Encyclopedia of Condensed Matter Physics (2nd
  Ed.)}}},\ \bibinfo {editor} {edited by\ \bibinfo {editor} {\bibfnamefont
  {T.}~\bibnamefont {Chakraborty}}}\ (\bibinfo  {publisher} {Academic Press},\
  \bibinfo {address} {Oxford},\ \bibinfo {year} {2024})\ pp.\ \bibinfo {pages}
  {84--123}\BibitemShut {NoStop}%
\bibitem [{\citenamefont {{Dyachenko}}\ \emph {et~al.}(1992)\citenamefont
  {{Dyachenko}}, \citenamefont {{Newell}}, \citenamefont {{Pushkarev}},\ and\
  \citenamefont {{Zakharov}}}]{Dyachenko:1992}%
  \BibitemOpen
  \bibfield  {author} {\bibinfo {author} {\bibfnamefont {S.}~\bibnamefont
  {{Dyachenko}}}, \bibinfo {author} {\bibfnamefont {A.~C.}\ \bibnamefont
  {{Newell}}}, \bibinfo {author} {\bibfnamefont {A.}~\bibnamefont
  {{Pushkarev}}},\ and\ \bibinfo {author} {\bibfnamefont {V.~E.}\ \bibnamefont
  {{Zakharov}}},\ }\bibfield  {title} {\bibinfo {title} {{Optical turbulence:
  weak turbulence, condensates and collapsing filaments in the nonlinear
  Schr{\"o}dinger equation}},\ }\href
  {https://doi.org/10.1016/0167-2789(92)90090-A} {\bibfield  {journal}
  {\bibinfo  {journal} {Physica D}\ }\textbf {\bibinfo {volume} {57}},\
  \bibinfo {pages} {96} (\bibinfo {year} {1992})}\BibitemShut {NoStop}%
\bibitem [{\citenamefont {{Christodoulou}}\ \emph {et~al.}(2021)\citenamefont
  {{Christodoulou}}, \citenamefont {{Ga{\l}ka}}, \citenamefont {{Dogra}},
  \citenamefont {{Lopes}}, \citenamefont {{Schmitt}},\ and\ \citenamefont
  {{Hadzibabic}}}]{Christodoulou:2021}%
  \BibitemOpen
  \bibfield  {author} {\bibinfo {author} {\bibfnamefont {P.}~\bibnamefont
  {{Christodoulou}}}, \bibinfo {author} {\bibfnamefont {M.}~\bibnamefont
  {{Ga{\l}ka}}}, \bibinfo {author} {\bibfnamefont {N.}~\bibnamefont {{Dogra}}},
  \bibinfo {author} {\bibfnamefont {R.}~\bibnamefont {{Lopes}}}, \bibinfo
  {author} {\bibfnamefont {J.}~\bibnamefont {{Schmitt}}},\ and\ \bibinfo
  {author} {\bibfnamefont {Z.}~\bibnamefont {{Hadzibabic}}},\ }\bibfield
  {title} {\bibinfo {title} {{Observation of first and second sound in a BKT
  superfluid}},\ }\href {https://doi.org/10.1038/s41586-021-03537-9} {\bibfield
   {journal} {\bibinfo  {journal} {Nature}\ }\textbf {\bibinfo {volume}
  {594}},\ \bibinfo {pages} {191} (\bibinfo {year} {2021})}\BibitemShut
  {NoStop}%
\bibitem [{\citenamefont {Etrych}\ \emph {et~al.}(2023)\citenamefont {Etrych},
  \citenamefont {Martirosyan}, \citenamefont {Cao}, \citenamefont {Glidden},
  \citenamefont {Dogra}, \citenamefont {Hutson}, \citenamefont {Hadzibabic},\
  and\ \citenamefont {Eigen}}]{Etrych:2023}%
  \BibitemOpen
  \bibfield  {author} {\bibinfo {author} {\bibfnamefont {J.}~\bibnamefont
  {Etrych}}, \bibinfo {author} {\bibfnamefont {G.}~\bibnamefont {Martirosyan}},
  \bibinfo {author} {\bibfnamefont {A.}~\bibnamefont {Cao}}, \bibinfo {author}
  {\bibfnamefont {J.~A.~P.}\ \bibnamefont {Glidden}}, \bibinfo {author}
  {\bibfnamefont {L.~H.}\ \bibnamefont {Dogra}}, \bibinfo {author}
  {\bibfnamefont {J.~M.}\ \bibnamefont {Hutson}}, \bibinfo {author}
  {\bibfnamefont {Z.}~\bibnamefont {Hadzibabic}},\ and\ \bibinfo {author}
  {\bibfnamefont {C.}~\bibnamefont {Eigen}},\ }\bibfield  {title} {\bibinfo
  {title} {{Pinpointing Feshbach resonances and testing Efimov universalities
  in $^{39}\mathrm{K}$}},\ }\href
  {https://doi.org/10.1103/PhysRevResearch.5.013174} {\bibfield  {journal}
  {\bibinfo  {journal} {Phys. Rev. Res.}\ }\textbf {\bibinfo {volume} {5}},\
  \bibinfo {pages} {013174} (\bibinfo {year} {2023})}\BibitemShut {NoStop}%
\bibitem [{\citenamefont {{Martirosyan}}\ \emph {et~al.}(2023)\citenamefont
  {{Martirosyan}}, \citenamefont {{Ho}}, \citenamefont {{Etrych}},
  \citenamefont {{Zhang}}, \citenamefont {{Cao}}, \citenamefont
  {{Hadzibabic}},\ and\ \citenamefont {{Eigen}}}]{Martirosyan:2023}%
  \BibitemOpen
  \bibfield  {author} {\bibinfo {author} {\bibfnamefont {G.}~\bibnamefont
  {{Martirosyan}}}, \bibinfo {author} {\bibfnamefont {C.~J.}\ \bibnamefont
  {{Ho}}}, \bibinfo {author} {\bibfnamefont {J.}~\bibnamefont {{Etrych}}},
  \bibinfo {author} {\bibfnamefont {Y.}~\bibnamefont {{Zhang}}}, \bibinfo
  {author} {\bibfnamefont {A.}~\bibnamefont {{Cao}}}, \bibinfo {author}
  {\bibfnamefont {Z.}~\bibnamefont {{Hadzibabic}}},\ and\ \bibinfo {author}
  {\bibfnamefont {C.}~\bibnamefont {{Eigen}}},\ }\bibfield  {title} {\bibinfo
  {title} {{Observation of subdiffusive dynamic scaling in a driven and
  disordered Bose gas}},\ }\href {https://doi.org/10.48550/arXiv.2304.06697}
  {\bibfield  {journal} {\bibinfo  {journal} {arXiv:2304.06697}\ } (\bibinfo
  {year} {2023})}\BibitemShut {NoStop}%
\bibitem [{\citenamefont {Zhang}\ \emph {et~al.}(2023)\citenamefont {Zhang},
  \citenamefont {Martirosyan}, \citenamefont {Ho}, \citenamefont {Etrych},
  \citenamefont {Eigen},\ and\ \citenamefont {Hadzibabic}}]{YZhang:2023}%
  \BibitemOpen
  \bibfield  {author} {\bibinfo {author} {\bibfnamefont {Y.}~\bibnamefont
  {Zhang}}, \bibinfo {author} {\bibfnamefont {G.}~\bibnamefont {Martirosyan}},
  \bibinfo {author} {\bibfnamefont {C.~J.}\ \bibnamefont {Ho}}, \bibinfo
  {author} {\bibfnamefont {J.}~\bibnamefont {Etrych}}, \bibinfo {author}
  {\bibfnamefont {C.}~\bibnamefont {Eigen}},\ and\ \bibinfo {author}
  {\bibfnamefont {Z.}~\bibnamefont {Hadzibabic}},\ }\bibfield  {title}
  {\bibinfo {title} {{Energy-space random walk in a driven disordered Bose
  gas}},\ }\href {https://doi.org/10.48550/arXiv.2309.12308} {\bibfield
  {journal} {\bibinfo  {journal} {arXiv:2309.12308}\ } (\bibinfo {year}
  {2023})}\BibitemShut {NoStop}%
\bibitem [{Note1()}]{Note1}%
  \BibitemOpen
  \bibinfo {note} {For our trap parameters, $a=30\protect \,a_0$ corresponds to
  the dimensionless 2D coupling strength $\protect \tilde {g} = 0.026$ and the
  Berezinskii-Kosterlitz-Thouless critical temperature is $T_{\protect \text
  {BKT}}= \SI {230}{\nK }$~\cite {Hadzibabic:2011}}\BibitemShut {NoStop}%
\bibitem [{\citenamefont {{Hadzibabic}}\ and\ \citenamefont
  {{Dalibard}}(2011)}]{Hadzibabic:2011}%
  \BibitemOpen
  \bibfield  {author} {\bibinfo {author} {\bibfnamefont {Z.}~\bibnamefont
  {{Hadzibabic}}}\ and\ \bibinfo {author} {\bibfnamefont {J.}~\bibnamefont
  {{Dalibard}}},\ }\bibfield  {title} {\bibinfo {title} {{Two-dimensional Bose
  fluids: An atomic physics perspective}},\ }\href
  {https://doi.org/10.1393/ncr/i2011-10066-3} {\bibfield  {journal} {\bibinfo
  {journal} {Riv. del Nuovo Cim.}\ }\textbf {\bibinfo {volume} {34}},\ \bibinfo
  {pages} {389}~(\bibinfo {year} {2011})}\BibitemShut {NoStop}%
\bibitem [{\citenamefont {Nazarenko}(2011)}]{Nazarenko:2011}%
  \BibitemOpen
  \bibfield  {author} {\bibinfo {author} {\bibfnamefont {S.}~\bibnamefont
  {Nazarenko}},\ }\href
  {https://link.springer.com/book/10.1007/978-3-642-15942-8} {\emph {\bibinfo
  {title} {Wave turbulence}}}\ (\bibinfo  {publisher} {Springer},\ \bibinfo
  {year} {2011})\BibitemShut {NoStop}%
\bibitem [{Note2()}]{Note2}%
  \BibitemOpen
  \bibinfo {note} {The corresponding prediction for weak three-wave turbulence,
  $\beta = -1/4$ and $\alpha =-1$, assumes a time-independent $n_0$ and is not
  applicable here. If the UV dynamics in our experiments were dominated by
  three-wave interactions, they would have to be even `faster' (with larger
  $|\beta |$), due to the growing $n_0$.}\BibitemShut {Stop}%
\bibitem [{Note20()}]{Note20}%
  \BibitemOpen
  \bibinfo {note} {{Given the value of $n_0$ at $t_1$, the universal-clock time
  $t_2$ must be such that $\Delta t$ later $n_0 = n_0(t_1) [(t_2 + \Delta
  t)/t_2]^{\alpha }$. For $t = t_1 + \Delta t$, this gives $n_0 (t) = n_0 (t_1)
  [(t - \protect \ensuremath {t^*}\protect \xspace )/(t_1 - \protect
  \ensuremath {t^*}\protect \xspace )]^{\alpha } \propto (t - \protect
  \ensuremath {t^*}\protect \xspace )^{\alpha }$}}\BibitemShut {NoStop}%
\bibitem [{Note21()}]{Note21}%
  \BibitemOpen
  \bibinfo {note} {Note that a similar mathematical structure appears in the
  analysis of finite-capacity turbulent cascades~\cite {Nazarenko:2011}, but
  with $t\rightarrow \protect \ensuremath {t^*}\protect \xspace \protect \! -
  t$; in that case $\protect \ensuremath {t^*}\protect \xspace $ denotes the
  time when the scaling ends.}\BibitemShut {Stop}%
\bibitem [{Note3()}]{Note3}%
  \BibitemOpen
  \bibinfo {note} {This is also hinted at by the fact that the three lines in
  Fig.~\ref {fig:2}D are parallel.}\BibitemShut {Stop}%
\bibitem [{\citenamefont {Gauthier}\ \emph {et~al.}(2019)\citenamefont
  {Gauthier}, \citenamefont {Reeves}, \citenamefont {Yu}, \citenamefont
  {Bradley}, \citenamefont {Baker}, \citenamefont {Bell}, \citenamefont
  {Rubinsztein-Dunlop}, \citenamefont {Davis},\ and\ \citenamefont
  {Neely}}]{Gauthier:2019}%
  \BibitemOpen
  \bibfield  {author} {\bibinfo {author} {\bibfnamefont {G.}~\bibnamefont
  {Gauthier}}, \bibinfo {author} {\bibfnamefont {M.~T.}\ \bibnamefont
  {Reeves}}, \bibinfo {author} {\bibfnamefont {X.}~\bibnamefont {Yu}}, \bibinfo
  {author} {\bibfnamefont {A.~S.}\ \bibnamefont {Bradley}}, \bibinfo {author}
  {\bibfnamefont {M.~A.}\ \bibnamefont {Baker}}, \bibinfo {author}
  {\bibfnamefont {T.~A.}\ \bibnamefont {Bell}}, \bibinfo {author}
  {\bibfnamefont {H.}~\bibnamefont {Rubinsztein-Dunlop}}, \bibinfo {author}
  {\bibfnamefont {M.~J.}\ \bibnamefont {Davis}},\ and\ \bibinfo {author}
  {\bibfnamefont {T.~W.}\ \bibnamefont {Neely}},\ }\bibfield  {title} {\bibinfo
  {title} {Giant vortex clusters in a two-dimensional quantum fluid},\ }\href
  {https://doi.org/10.1126/science.aat5718} {\bibfield  {journal} {\bibinfo
  {journal} {Science}\ }\textbf {\bibinfo {volume} {364}},\ \bibinfo {pages}
  {1264} (\bibinfo {year} {2019})}\BibitemShut {NoStop}%
\bibitem [{\citenamefont {Johnstone}\ \emph {et~al.}(2019)\citenamefont
  {Johnstone}, \citenamefont {Groszek}, \citenamefont {Starkey}, \citenamefont
  {Billington}, \citenamefont {Simula},\ and\ \citenamefont
  {Helmerson}}]{Johnstone:2019}%
  \BibitemOpen
  \bibfield  {author} {\bibinfo {author} {\bibfnamefont {S.~P.}\ \bibnamefont
  {Johnstone}}, \bibinfo {author} {\bibfnamefont {A.~J.}\ \bibnamefont
  {Groszek}}, \bibinfo {author} {\bibfnamefont {P.~T.}\ \bibnamefont
  {Starkey}}, \bibinfo {author} {\bibfnamefont {C.~J.}\ \bibnamefont
  {Billington}}, \bibinfo {author} {\bibfnamefont {T.~P.}\ \bibnamefont
  {Simula}},\ and\ \bibinfo {author} {\bibfnamefont {K.}~\bibnamefont
  {Helmerson}},\ }\bibfield  {title} {\bibinfo {title} {Evolution of
  large-scale flow from turbulence in a two-dimensional superfluid},\ }\href
  {https://doi.org/10.1126/science.aat5793} {\bibfield  {journal} {\bibinfo
  {journal} {Science}\ }\textbf {\bibinfo {volume} {364}},\ \bibinfo {pages}
  {1267} (\bibinfo {year} {2019})}\BibitemShut {NoStop}%
\bibitem [{\citenamefont {{Groszek}}\ \emph {et~al.}(2020)\citenamefont
  {{Groszek}}, \citenamefont {{Davis}},\ and\ \citenamefont
  {{Simula}}}]{Groszek:2020}%
  \BibitemOpen
  \bibfield  {author} {\bibinfo {author} {\bibfnamefont {A.~J.}\ \bibnamefont
  {{Groszek}}}, \bibinfo {author} {\bibfnamefont {M.~J.}\ \bibnamefont
  {{Davis}}},\ and\ \bibinfo {author} {\bibfnamefont {T.~P.}\ \bibnamefont
  {{Simula}}},\ }\bibfield  {title} {\bibinfo {title} {{Decaying quantum
  turbulence in a two-dimensional Bose-Einstein condensate at finite
  temperature}},\ }\href {https://doi.org/10.21468/SciPostPhys.8.3.039}
  {\bibfield  {journal} {\bibinfo  {journal} {SciPost Phys.}\ }\textbf
  {\bibinfo {volume} {8}},\ \bibinfo {eid} {039} (\bibinfo {year}
  {2020})}\BibitemShut {NoStop}%
\bibitem [{\citenamefont {Bland}\ \emph {et~al.}(2022)\citenamefont {Bland},
  \citenamefont {Poli}, \citenamefont {Politi}, \citenamefont {Klaus},
  \citenamefont {Norcia}, \citenamefont {Ferlaino}, \citenamefont {Santos},\
  and\ \citenamefont {Bisset}}]{Bland:2022}%
  \BibitemOpen
  \bibfield  {author} {\bibinfo {author} {\bibfnamefont {T.}~\bibnamefont
  {Bland}}, \bibinfo {author} {\bibfnamefont {E.}~\bibnamefont {Poli}},
  \bibinfo {author} {\bibfnamefont {C.}~\bibnamefont {Politi}}, \bibinfo
  {author} {\bibfnamefont {L.}~\bibnamefont {Klaus}}, \bibinfo {author}
  {\bibfnamefont {M.~A.}\ \bibnamefont {Norcia}}, \bibinfo {author}
  {\bibfnamefont {F.}~\bibnamefont {Ferlaino}}, \bibinfo {author}
  {\bibfnamefont {L.}~\bibnamefont {Santos}},\ and\ \bibinfo {author}
  {\bibfnamefont {R.~N.}\ \bibnamefont {Bisset}},\ }\bibfield  {title}
  {\bibinfo {title} {Two-dimensional supersolid formation in dipolar
  condensates},\ }\href {https://doi.org/10.1103/PhysRevLett.128.195302}
  {\bibfield  {journal} {\bibinfo  {journal} {Phys. Rev. Lett.}\ }\textbf
  {\bibinfo {volume} {128}},\ \bibinfo {pages} {195302} (\bibinfo {year}
  {2022})}\BibitemShut {NoStop}%
\end{thebibliography}
\end{document}